\documentclass[12pt]{article}
\usepackage[dvips]{color,graphicx}
\usepackage{amsmath,amssymb}
\usepackage{verbatim}
\DeclareGraphicsExtensions{.eps}

\newcommand{\ls}{\ensuremath{l_s}} 

\def\p{\partial}

\newcommand{\tr}{\mathop{\rm Tr}}

\def\expec#1{\langle #1 \rangle}

\newcommand{\cF}{{\mathcal{F}}}

\newcommand{\cL}{\mathcal{L}}

\newcommand{\cN}{{\mathcal{N}}}
\newcommand{\cO}{{\mathcal{O}}}

\newcommand{\bS}{{\mathbf{S}}}

\newcommand{\tret}{{t_{\mbox{\scriptsize ret}}}}

\begin{document}

\begin{titlepage}

\begin{flushright}
UTTG-22-11\\
TCC-024-11
\end{flushright}

\begin{center} \Large \bf Holographic Lessons for Quark Dynamics
\end{center}

\begin{center}
Mariano Chernicoff$^{1}$,
J.~Antonio Garc\'{\i}a$^{2}$,
Alberto G\"uijosa$^{2}$
and Juan F.~Pedraza$^{3}$

\vspace{0.2cm}
$^{1}$
Departament de F{\'\i}sica Fonamental and Institut de Ci\`encies del Cosmos,\\
Universitat de Barcelona,\\
Marti i Franqu\`es 1, E-08028 Barcelona, Spain\\
 \vspace{0.2cm}
$^{2}\,$Departamento de F\'{\i}sica de Altas Energ\'{\i}as, Instituto de Ciencias Nucleares, \\ Universidad Nacional Aut\'onoma de
M\'exico,\\ Apartado Postal 70-543, M\'exico D.F. 04510, M\'exico\\
 \vspace{0.2cm}
$^{3}\,$Theory Group, Department of Physics and Texas Cosmology Center, \\
University of Texas, 1 University Station C1608, Austin, TX 78712, USA\\
\vspace{0.2cm}
{\tt mchernicoff@ub.edu, garcia@nucleares.unam.mx, alberto@nucleares.unam.mx, jpedraza@physics.utexas.edu}
\vspace{0.2cm}
\end{center}

\begin{center}
{\bf Abstract}
\end{center}
\noindent
 We give a brief overview of recent results obtained through the gauge/gravity correspondence, concerning the propagation of a heavy quark in strongly-coupled conformal field theories (such as $\mathcal{N}=4$ super-Yang-Mills), both at zero and finite temperature. In the vacuum, we discuss energy loss, radiation damping, signal propagation and radiation-induced fluctuations. In the presence of a thermal plasma, our emphasis is on early-time energy loss, screening and quark-antiquark evolution after pair creation. Throughout, quark dynamics is seen to be efficiently encapsulated in the usual string worldsheet dynamics.
\vspace{0.2in}
\smallskip
\end{titlepage}
\setcounter{footnote}{0}


\tableofcontents

\vspace{0.5cm}

\section{Introduction}

A substantial amount of evidence from heavy ion collision experiments at RHIC and, recently, LHC, supports the detection of the long-sought quark-gluon plasma (QGP), a hot and dense phase of deconfined strongly-interacting matter \cite{qgprev}. Energetic partons serve as important probes of this thermal medium, and there exists an enormous `jet-quenching' literature dedicated to analyzing the manner in which the plasma damps their motion, and is in turn disturbed by their passage \cite{energylossrev}. Given the experimental indications that, in the range of accessible temperatures (of order a few times the deconfinement temperature), the QGP is strongly-coupled, perturbative QCD is believed to be inadequate for at least some aspects of the relevant calculations, creating a demand for new theoretical tools. In the past five years, interesting steps have been taken towards meeting this demand via the AdS/CFT, or, more generally, gauge/gravity or holographic correspondence \cite
 {malda,magoo}, starting with the seminal works \cite{hkkky,gubser,ct,liuqhat} and continuing with a large body of work that has been reviewed in \cite{clmrw}.

The gauge/gravity correspondence is by now well-established as a tool that grants access (often analytically) to a large class of strongly-coupled non-Abelian gauge theories, via a drastic and surprising rewriting in terms of string-theoretic (frequently just supergravity) degrees of freedom living on a curved higher-dimensional geometry. The gauge/gravity catalog includes some theories that display characteristic QCD-like physics such as confinement and chiral-symmetry breaking, but not, to date, QCD itself.
Thus, at our present stage of knowledge, contact with the real-world QGP produced at RHIC or LHC will be feasible only if QCD can be reasonably well approximated by at least
one of the gauge theories whose dual description is known. In the past few years, encouraging signs in this direction have emerged even for the most rudimentary
example \cite{malda}, $SU(N_c)$ maximally-supersymmetric ($\mathcal{N}=4$) Yang-Mills theory (MSYM), a conformal field theory (CFT) that at zero temperature is completely unlike QCD (and in particular, has a coupling that does not run), but at finite temperature is in various respects analogous to
deconfined QCD.  Besides the gauge field, MSYM contains 6 real scalar fields and 4 Weyl fermions, all in the adjoint representation of the gauge group.

Following the very exciting findings on the shear
viscosity in holographic thermal plasmas \cite{eta} (which in turn led to calculation of many other transport coefficients \cite{etarev}),
analyses of energy loss have been conducted considering various types of partonic probes of the MSYM plasma, including quarks \cite{hkkky,gubser,ct,liu}, mesons \cite{sonnenschein,liuwind,dragqqbar,dusling}, baryons \cite{draggluon,liu5,krishnan}, gluons \cite{draggluon,gubsergluon} $k$-quarks \cite{draggluon} and various types of defects \cite{jankarch}.
Energy loss studies have also been carried out via the gauge/gravity correspondence in many other theories, including some with properties more akin to those of QCD--- see, e.g., \cite{bigazzi,otherdragforce} and references therein\footnote{See also \cite{panero} for an interesting comparison between lattice results and its holographic counterpart.}. Other examples of drag force calculations may be found in the reviews \cite{clmrw}.
 While we are probably still far from achieving firm contact between experimental data and first-principles gauge/gravity calculations, the correspondence has already succeeded in enhancing our general intuition on the behavior of gauge theories at strong coupling,
 and has additionally provided useful suggestions for phenomenological models of the QGP (see, e.g., \cite{horowitz,bngt,cm,cdm,kharzeev}).
In the present paper, we will review a number of recent results related to the evolution of a quark in a strongly-coupled gluonic plasma.

\section{Quarks, with Strings Attached}

\subsection{Basic setup}\label{setupsec}

For our analysis we need two ingredients: a strongly-coupled thermal non-Abelian plasma, and a quark that traverses it. Via the gauge/gravity correspondence, a plasma in a $d$-dimensional CFT is known to be described in dual language by a Schwarzschild black hole (more precisely, black brane) in $(d+1)$-dimensional asymptotically anti-de Sitter (AdS) space. For concreteness, we will frame our discussion in the context of $SU(N_c)$
MSYM in $3+1$ dimensions, with coupling $g_{YM}$ and temperature $T$, which is equivalent to Type IIB string theory on the (AdS-Schwarzschild)$_5\times\bS^5$ geometry
\begin{eqnarray}\label{metric}
ds^2&=&G_{mn}dx^m dx^n={R^2\over z^2}\left(
-hdt^2+d\vec{x}^2+{dz^2 \over h}\right)+R^2 d\Omega_5~, \\
h&=&1-\frac{z^{4}}{z_h^4}~, \qquad {R^4\over \ls^4}=g_{YM}^2 N_c\equiv\lambda~, \qquad
z_h={1\over \pi T}~,\nonumber
\end{eqnarray}
(with a constant dilaton and $N_c$ units of Ramond-Ramond five-form flux through the five-sphere), where $\ls$ denotes the string length, and $z_h$ is the location of the event horizon. The
radial direction $z$ is mapped holographically into a variable length scale in the gauge
theory, in such a way that $z\to 0$ and $z\to\infty$ are respectively the ultraviolet and infrared limits. The directions $x^{\mu}\equiv(t,\vec{x})$ are parallel to the AdS boundary $z=0$ and are
directly identified with the gauge theory directions. The five-sphere coordinates are associated with the global $SU(4)$ internal (R-) symmetry of MSYM. They will play no role in our discussion, so all results below will hold equally well in the more general case where the $\bS^5$ is replaced by a different compact five-dimensional space $\mathbf{X}_5$, which corresponds to replacing $\cN=4$ SYM with a different $(3+1)$-dimensional CFT.

 Notice that the field theory temperature is identified with the Hawking temperature $T_H=1/\pi z_h$ of the black hole. In the limit of
vanishing temperature ($z_h\to\infty$), we are left
in (\ref{metric}) with a pure AdS geometry, which is dual to the (symmetry-preserving) vacuum of MSYM.  The closed string sector describing (small or large) fluctuations on top of AdS fully captures the nonperturbative gluonic ($+$ adjoint scalar and fermionic) physics. The generic (AdS-Schwarzschild)$_5\times\bS^5$ geometry (\ref{metric}) is a special case of such a fluctuation, and is dual to a thermal ensemble in MSYM. The string theory description is under calculational control only for small string coupling and low curvatures, which translates into $N_c\gg 1$,  $\lambda\gg 1$, i.e., a large number of colors and strong ('t~Hooft) coupling.

It is also known that one can add $N_f$ flavors of matter in the \emph{fundamental}
representation of the $SU(N_c)$ gauge group
by introducing in the string theory setup an open string sector associated with a stack of $N_f$ D7-branes \cite{kk}.
 We will refer to these degrees of freedom
as `quarks,' even though, being $\cN=2$ supersymmetric,
they include
both spin $1/2$ and spin $0$ fields. For $\lambda N_f \ll N_c$,
 we are allowed to neglect the backreaction of the D7-branes on the geometry \cite{bigazzi,unquenched};
 in the gauge theory this corresponds to working in a `quenched' approximation
that ignores quark loops.
 These branes cover the four gauge theory directions $x^{\mu}$, and are spread
along the radial AdS direction from $z=0$ to $z=z_m$. The D7-brane
parameter $z_m$ is related to the Lagrangian mass $m\gg \sqrt{\lambda}T$ of the
quark through \cite{hkkky}
\begin{equation}\label{zm} {1\over z_m}={2\pi
m\over\sqrt{\lambda}}\left[1+{1\over 8}\left(\sqrt{\lambda} T\over 2 m\right)^4-{5\over
128}\left(\sqrt{\lambda} T\over 2 m\right)^8+\cO\left(\left(\sqrt{\lambda} T\over 2
m\right)^{12}\right)\right]~,
\end{equation}
which simplifies to
\begin{equation}\label{zmnoplasma}
z_m={\sqrt{\lambda}\over 2\pi m}~
\end{equation}
at vanishing temperature. For applications of this formalism to phenomenology, we must choose values of the mass
parameter $z_m$ based on the charm and bottom quark masses, $m\simeq 1.4,4.8$ GeV. Following
\cite{gubsercompare}, at the temperatures relevant to RHIC this translates into $z_m/z_h\sim 0.16-0.40$ for charm and
$z_m/z_h\sim 0.046-0.11$ for bottom.

An isolated heavy quark corresponds to a string extending up from
the horizon at $z=z_h$ to a location $z=z_m$ where it ends on the  stack of $N_f$ D7-branes.
The string dynamics is governed by the Nambu-Goto action
\begin{equation}\label{nambugoto}
S_{NG}=-{1\over 2\pi\ls^2}\int
d^2\sigma\,\sqrt{-\det{g_{ab}}}\equiv \int
d^2\sigma\,\cL_{{ NG}}~, 
\end{equation}
where $g_{ab}\equiv\p_a X^M\p_b X^N G_{MN}(X)$ ($a,b=0,1$) denotes
the induced metric on the worldsheet.
We can exert an external force $\vec{F}$ on the string endpoint by turning on an electric field $F_{0i}=F_i$ on the D7-branes. This amounts to adding to (\ref{nambugoto}) the
usual minimal coupling, which in terms of the endpoint/quark worldline
$x^{\mu}(\tau)\equiv X^{\mu}(\tau,z_m)$ reads
\begin{equation}\label{externalforce}
S_{{ F}}=\int
d\tau\,A_{\mu}(x(\tau))\,{dx^{\mu}(\tau)\over d\tau}~.
\end{equation}
Variation of  $S_{{ NG}}+S_{{ F}}$ implies the standard Nambu-Goto equation of motion for all interior points of the string, plus the boundary condition
\begin{equation}\label{stringbc}
\Pi^{z}_{\mu}(\tau)|_{z=z_m}=\cF_{\mu}(\tau)\quad\forall~\tau~,
\end{equation}
where $\Pi^{z}_{\mu}\equiv {\p\cL_{{ NG}}}/{\p(\p_z X^{\mu})}$
is the worldsheet current associated with spacetime momentum, and
$\cF_{\mu}=F_{\mu\nu}\p_{\tau}x^{\nu}
=(-\gamma\vec{F}\cdot\vec{v},\gamma\vec{F})$
the Lorentz four-force.

Notice that the string is being described (as is customary) in first-quantized
language, and, as long as it is sufficiently heavy, we are allowed to treat it semiclassically.
In gauge theory language, then,
we are coupling a first-quantized quark to the gluonic ($+$ other MSYM) field(s), and then carrying out the full path integral over the strongly-coupled field(s) (the result of which is codified by the AdS spacetime), but, for the time being, treating the path integral over the quark trajectory $x^{\mu}(\tau)$ in a saddle-point approximation. (The effect of quantum fluctuations about the classical string configuration will be discussed in Sections \ref{latetimesec} and \ref{unruhsec}.)

 A crucial property of the gauge/gravity dictionary is that it identifies the endpoint(s) of the string as being dual to the quark (or antiquark), while the body of the string codifies the profile of the (near and radiation) gluonic ($+$ other MSYM) field(s) set up by the quark. In other words, this dictionary teaches us that the usual `QCD' string exists even for non-confining theories, is infinitesimally thin, and actually lives in a curved 5 ($+$ 5)-dimensional geometry. The gluonic profile can be mapped out explicitly by computing one-point functions of local operators ($\expec{\tr F^2(x)}$,$\expec{T_{\mu\nu}(x)}$, $\ldots$) in the presence of the quark, which, via the GKPW recipe \cite{gkpw}, requires a determination of the near-boundary profile of the closed string fields ($\phi(x,z)$, $h_{\mu\nu}(x,z)$, $\ldots$) generated by the macroscopic string.

It is important to keep in mind that the quark described by this string is not `bare' but `composite' or `dressed'. In an interacting field theory, a particle of finite mass can only be regarded as strictly pointlike at zeroth-order in a perturbative description. In the nonperturbative framework made available to us by the gauge/gravity duality, a quark with finite mass automatically acquires a finite size. This can be seen most clearly by working out the vacuum expectation value of the gluonic field surrounding a static quark located at the origin \cite{martinfsq}. For $m\to\infty$
($z_m\to 0$), this is just the Coulombic field expected (by conformal invariance) for a pointlike
charge. For finite $m$, the profile is still Coulombic far away from the origin, but in fact becomes
non-singular at the location of the quark. The characteristic thickness of the implied non-Abelian charge distribution
 is precisely the length scale $z_m$ defined in (\ref{zm}). This is then the size of
the gluonic cloud that surrounds the quark, or in other words, the analog of the Compton wavelength for our non-Abelian source.

 Below we will also be interested in the description of a quark-antiquark pair. The IIB strings we have introduced above are oriented, and a state with two oppositely oriented purely radial strings would correspond to a quark and antiquark that are merely superposed. With such boundary conditions, however, the configuration with lowest energy is given by a \emph{single} $\cup$-shaped string that has both of its endpoints at $z=z_m$. This can again be verified by computing the corresponding gluonic profile in vacuum \cite{cg,linshuryak}, to verify that the falloff is more rapid than Coulombic, as expected for an overall color-neutral source.

\subsection{Late-time energy loss, Brownian motion and limiting velocity}
\label{latetimesec}

 Combining the two preceding ingredients, we know that a heavy quark in a strongly-coupled MSYM plasma is described by a string on the Schwarzschild-AdS geometry (\ref{metric}). As the string endpoint moves, its body lags behind it, exerting a drag on the tip that is the gravity-side realization of the damping force exerted by the plasma on the quark, as studied in \cite{hkkky,gubser}.
 For a restricted type of quark trajectories,   this mechanism leads to energy loss at a rate
\begin{equation}\label{elossgubser}
\frac{dE_{\mbox{\scriptsize rad}}}{dt}=\frac{\pi}{2}\sqrt{\lambda}T^2
\frac{v^2}{\sqrt{1-v^2}}~.
\end{equation}
Various works have explored the way in which these heavy quark results are modified for heavy sources of the gluonic field in color representations other than the fundamental (including the adjoint) \cite{draggluon}, as well as for \emph{light} quarks and gluons \cite{lightpartonenergyloss,iancujet}.

In the context of weakly-coupled QCD, it is known that energy loss at high parton velocity is dominated by the strong-interaction analog of bremsstrahlung, i.e., medium-induced gluonic radiation, while for low velocities collisional loss becomes important \cite{energylossrev}. In the strongly-coupled MSYM setup made available to us by AdS/CFT, the perturbative terminology is no longer adequate. The flow of energy/momentum along the body of the trailing string corresponds to energy/momentum transported away from the quark via the gluonic field generated jointly by the quark and the plasma, in a pattern that has been meticulously studied in a large body of work that began with \cite{gluonicprofile}, has been reviewed in \cite{gluonicprofilerev} and includes the recent additions \cite{liusynchrotron,iancu1,veronika,iancu2,trfsq,baier}.
 For ease of language, throughout this paper we will continue to speak of radiation, even though this concept is not really appropriate within an infinite thermal medium.

We should stress that, within the framework of \cite{hkkky,gubser}, adopted also in this paper, the entire energy loss calculation is carried out inside the MSYM theory, whose coupling does not run and is taken to be large. For quarks that are extremely e\-ner\-ge\-tic, the asymptotic freedom of QCD would lead one to expect deviations between this scenario and the real-world QGP, and seek a hybrid approach describing hard/perturbative emission of a gluon that subsequently propagates through a strongly-coupled medium, as advocated in \cite{liuqhat,liuwilson}.\footnote{The approach of \cite{liuqhat,liuwilson} involves the extraction from a lightlike Wilson loop of a transport (`jet-quenching') parameter $\hat{q}$, which characterizes the medium and whose value at strong coupling curiously differs from the customary definition of $\hat{q}$ as the average transverse momentum picked up per unit distance traveled \cite{gubserqhat,ctqhat}. Some doubts about the procedure proposed in
  \cite{liuqhat,liuwilson} to calculate $\hat{q}$ on the gravity side of the correspondence had been raised by, e.g., \cite{dragqqbar,argyres2,argyres3}, but these concerns appear to have been resolved by a recent correction to the procedure \cite{elr}, which happens to leave the result unchanged.} It is debatable, however, which of these two scenarios is more appropriate at the not extremely relativistic heavy quark energies achieved at RHIC. At the very least, the calculations discussed in this paper have a direct interpretation in terms of energy loss in a thermal plasma of strongly-coupled gluons and exotic matter (`XGP') of MSYM (or other CFTs), which is an interesting theoretical question in its own right.

In \cite{hkkky,gubser}, the rate (\ref{elossgubser}) was deduced for an isolated quark
 in the stationary or late-time regimes, i.e., assuming that the quark either moves
at constant velocity, under the influence of an external force which precisely cancels the drag force exerted by the plasma, or has
decelerated solely under the action of this drag force for a long period of time, and is about to come to rest. In the actual experimental setup, however, the configuration
is neither stationary nor asymptotic:
the quark is not externally forced, and slows down under the influence of the plasma, whose finite spatial and temporal extent imply that the
late-time regime is not generically accessible.\footnote{This latter point has been emphasized from the phenomenological perspective in \cite{peigne1}.} Moreover, the real-world quark is not isolated, but is created within the plasma together with an antiquark.
In Section \ref{qplasmasec} we will explore to what extent the rate of energy loss is affected by these issues. Also, as we will seen in Section \ref{qnoplasmasec}, the above trailing string mechanism is in fact much more general, and accounts even for the radiative damping expected in vacuum \cite{lorentzdirac,damping}. That is, irrespective
of whether a \emph{spacetime} black hole is present or not, the body of the string plays the role of an energy sink, as befits its identification as the embodiment of the gluonic degrees of freedom. On the other hand, energy loss via the string does turn out to be closely
associated with the appearance of a \emph{worldsheet} horizon,
  as noticed initially in \cite{gubserqhat,ctqhat} at finite temperature and in
  \cite{dragtime} (see also \cite{dominguez,xiao}) for the zero temperature case.

 The appearance of a black hole on the string worldsheet is crucial in particular to reproduce the expected Brownian motion of the dual quark in the hot medium.   This connection was first worked out in detail in \cite{rangamani,sonteaney}, for a static string on the Schwarzschild-AdS$_{d+1}$ geometry, which as we know from Section \ref{setupsec} is dual to a static quark in a thermal bath of the CFT. At finite $\lambda$, one must take into account small perturbations about the average string embedding. These are described by free scalar fields propagating on the worldsheet black hole geometry, and upon quantization, they are excited by Hawking radiation. This leads to Brownian motion of the endpoint/quark, whose detailed form is captured by a generalized Langevin equation. The authors of \cite{rangamani} reached this conclusion in arbitrary dimension by assuming that the state of the quantized embedding fields is the usual Hartle-Hawking (or Kruskal) vacuum, which describes
 the black hole in equilibrium with its own thermal radiation. The authors of \cite{sonteaney} focused on the case $d=4$ and followed a different but equivalent route, employing the dual relation between the Kruskal extension of the Schwarzschild-AdS geometry and the CFT Schwinger-Keldysh formalism \cite{maldaeternal,herzogson,ct}, together with the known connection between the latter and the generalized Langevin equation. These calculations were later generalized and elaborated on in \cite{iancu,giecold,casalderrey,deboer}.

Another very interesting prediction of AdS/CFT in this thermal context is the existence of a subluminal limiting velocity
\begin{equation} \label{vm}
v_m\equiv \sqrt{1-\frac{z_m^4}{z_h^4}}\simeq 1-\left(\sqrt{\lambda}\frac{T}{2m}\right)^4
\end{equation}
for the quark traversing the plasma. This follows simply from the fact that the quark velocity is dual to the \emph{coordinate} velocity of the string endpoint, and $v=v_m$ corresponds to a \emph{proper} velocity at $z=z_m$ equal to that of light \cite{argyres1}. A more general bound involving the external force $\vec{F}$ can be derived by requiring the Nambu-Goto square root to remain real \cite{dragtime}. This same restriction on the velocity can be seen to arise from microscopic calculations of the meson spectrum \cite{mateosthermo,liu4}. Its validity for isolated quarks has been emphasized in \cite{argyres3,dragtime}. The existence of this limiting velocity might have interesting phenomenological consequences, such as the photon peak predicted in \cite{cm} or the Cherenkov emission of mesons analyzed in \cite{cdm}.

\section{Isolated Quark at Zero Temperature}
\label{qnoplasmasec}

\subsection{Energy loss in vacuum}
\label{energylosssec}

To orient ourselves, we will first inquire into the rate of energy loss for an acce\-le\-ra\-ting quark in vacuum, whose dual description involves the string moving on pure AdS spacetime. In this case we will have full analytic control over the system, which will allow us to develop some intuition on the problem at hand. Besides being of theoretical interest in its own right, our vacuum analysis will provide a useful benchmark against which the finite-temperature results can be compared.

A quark that accelerates in vacuum would be expected to emit chromoelectromagnetic
radiation. The first definite characterization of the radiation rate off an accelerating quark by means of AdS/CFT was
worked out in an important paper by Mikhailov \cite{mikhailov}. Remarkably, this author was able to solve the highly nonlinear
equation of motion for a string on AdS$_5$ that follows from (\ref{nambugoto}), for an \emph{arbitrary} timelike trajectory of
the string endpoint dual to an infinitely massive quark. In terms of the coordinates used in
(\ref{metric}) (where for now $h=1$), his solution is
\begin{equation}\label{mikhsol}
X^{\mu}(\tau,z)=z{dx^{\mu}(\tau)\over d\tau}+x^{\mu}(\tau)~,
\end{equation}
with $\mu=0,1,2,3$, and $x^{\mu}(\tau)$ the worldline of the string endpoint at the AdS
boundary--- or, equivalently, the worldline of the dual quark---
parametrized by the proper time $\tau$ defined through
$\eta_{\mu\nu}(dx^{\mu}/d\tau)(dx^{\nu}/d\tau)=-1$.

{}From the structure of (\ref{mikhsol}) we see that the behavior of the string segment located at a given time $t$ and radial depth $z$ (which according to our discussion in the previous section codifies the behavior of the gluonic field at the length scale $z$ \cite{uvir}) is completely determined by the behavior of the quark/string endpoint at the earlier, \emph{retarded} time $\tret(t,z)$ defined by
\begin{equation}\label{tret}
t=z{1\over\sqrt{1-\vec{v}(\tret)^{\,2}}}+\tret~,
\end{equation}
where the quark worldline has been parametrized by $x^0(\tau)$
instead of $\tau$, and $\vec{v}\equiv
d\vec{x}/dx^0$.
The definition (\ref{tret}) can be shown to imply that $\tret$ is obtained by projecting back to the AdS boundary along a (fixed $\tau$) curve that is null on the string worldsheet \cite{mikhailov}, in analogy with the  Lienard-Wiechert story in classical electrodynamics. The embedding (\ref{mikhsol}) thus describes a wave on the string that is  \emph{purely outgoing}: it is generated at the string endpoint and then descends along the body of the string, moving into the AdS bulk. This is why, among the infinite number of extremal string embeddings that satisfy the boundary condition associated with the given quark trajectory, the profile (\ref{mikhsol}) is  the one that is of physical interest for us: it is dual to a configuration where waves in the gluonic field move out from the quark to infinity.

Working in the static gauge $\sigma^0=t$, $\sigma^1=z$, Mikhailov was able to reexpress the total energy of the string embedding (\ref{mikhsol}) (via a
change of integration variable $z\to\tret$) as a local functional of the quark trajectory,
\begin{equation}\label{emikh}
E(t)={\sqrt{\lambda}\over 2\pi}\int^t_{-\infty}d\tret
\frac{\vec{a}^{\,2}-\left[\vec{v}\times\vec{a}\right]^2}{\left(1-\vec{v}^{\,2}\right)^3}
+E_q(\vec{v}(t))~,
\end{equation}
where of course $\vec{a}\equiv d\vec{v}/dx^0$.
The second term in the above equation arises
from a total derivative that was not explicitly written down by Mikhailov, but can easily be
worked out to be \cite{dragtime}
\begin{equation}\label{edr}
E_q(\vec{v})={\sqrt{\lambda}\over
2\pi}\left.\left({1\over\sqrt{1-\vec{v}^{\,2}}}{1\over
z}\right)\right|^{z_m=0}_{\infty}=\gamma m~,
\end{equation}
which gives the expected Lorentz-invariant dispersion relation for
the quark. The energy split achieved in (\ref{emikh}) therefore
admits a clear and pleasant physical interpretation: $E_q$ (associated only
with information of the string endpoint) is the
intrinsic energy of the quark at time $t$, and the integral over
$\tret$ (associated with the body of the string) encodes the accumulated energy \emph{lost} by the quark to its gluonic field
over all times prior to $t$.\footnote{This energy split has been extended to $k$-quarks in the recent work \cite{fiol}.} Completely analogous statements can be derived for the spatial momentum. No less remarkable is the fact that
the rate of energy loss for the quark in this strongly-coupled
non-Abelian theory is found to be in precise agreement with the
standard Lienard formula from classical electrodynamics!
The AdS/CFT correspondence thus teaches us that, in this very unfamiliar nonlinear setting, the energy loss turns out to depend only locally on the quark worldline. This feature has been argued in \cite{kharzeev} to lead to an upper bound for the energy of a quark at finite temperature.  In the next subsection we will see that both the rate of radiation and the dispersion relation are modified when the quark has a finite mass, and is therefore not pointlike.

\subsection{Radiation damping} \label{dampingsec}

The radiation emitted by an accelerated charge inevitably backreacts on the charge. One effect, present already at the classical level, is a reaction force on the charge, that tends to damp its motion. And if the system is quantized,
 one additionally expects the emission of radiation to induce stochastic fluctuations of the charge's trajectory. The discovery of the gauge/gravity duality has opened the possibility of exploring these effects in the previously uncharted terrain of strongly-coupled non-Abelian gauge theories. We will discuss here radiation damping, and postpone the analysis of fluctuations until Section \ref{unruhsec}.

  In the context of
classical electrodynamics, and for an electron modeled as a
vanishingly small charge distribution, the effect of radiation damping is incorporated in the classic  (Abraham-)Lorentz-Dirac equation \cite{dirac},
\begin{equation}\label{ald}
m\left({d^2 x^{\mu}\over d\tau^2}-t_e\left[
{d^3 x^{\mu}\over d\tau^3}-{1\over c^2}{d^2 x_{\nu}\over d\tau^2}\,
{d^2 x^{\nu}\over d\tau^2}{dx^{\mu}\over d\tau}\right]\right)=\cF^{\mu}~,
\end{equation}
with
$\tau$ the proper time, $\cF^{\mu}\equiv\gamma(\vec{F}\cdot\vec{v}/c,\vec{F})$
the external 4-force, and $t_e\equiv 2e^2/3mc^3$ a characteristic timescale set by the classical electron radius. The second term within the square brackets
is the negative of the rate at which 4-momentum is carried away from the charge by radiation (as given by the covariant Lienard formula), so it is only this term that can properly be called radiation reaction. The first term within the square brackets, usually called the Schott term, is known to arise from the effect of the charge's `near' (as opposed to radiation) field \cite{teitelboim,rohrlich}.
The appearance of a third-order term in (\ref{ald}) leads to unphysical
behavior, including pre-accelerating and self-accelerating (or `runaway')  solutions. These deficiencies are known to
originate from the assumption that the charge is pointlike.

Analysis of radiation damping in the quantum non-Abelian case using traditional me\-thods would be a serious
challenge, but the AdS/CFT correspondence
 allows us to address it rather easily in certain
strongly-coupled non-Abelian gauge theories \cite{lorentzdirac,damping}.
 By Lorentz invariance, given the description of the static quark, we know that a quark moving at constant velocity corresponds to a purely radial string, moving as a rigid vertical rod. When the quark accelerates, the body of the string will trail behind the endpoint, and will therefore exert a force on the latter. Remembering that the body of the string codifies the SYM fields sourced by the quark, we know that in the gauge theory this force is interpreted as the backreaction of the gluonic field on the quark. In other words, in the AdS/CFT context the quark always has a `tail', and it is this tail that is responsible for the damping effect we are after.
This is of course the same mechanism that yields the drag force exerted on the quark by a thermal plasma \cite{hkkky,gubser}, reviewed in the previous section. The analysis of \cite{lorentzdirac,damping}  makes it clear that the damping effect is equally present in the gauge theory vacuum.

To obtain a noticeable damping effect, we must extend the analysis of the preceding subsection to the more interesting case with $z_m>0$, where the quark has a finite mass. As we have
emphasized in Section \ref{setupsec}, in this case our non-Abelian source is no longer pointlike but has size $z_m$. On the string theory side, the string endpoint is now at $z=z_m$, and we must again require it to follow the given quark trajectory, $x^{\mu}(\tau)$. As before, this condition by itself does not pick out a unique string embedding. Just like we discussed for the infinitely massive case, we additionally require the solution to be retarded, in order to focus on the gluonic field causally set up by the quark. As in \cite{dragtime}, we can inherit this structure by truncating a suitably selected retarded Mikhailov solution.
The embeddings of interest to us can thus be regarded as the $z\ge z_m$ portions of the solutions (\ref{mikhsol}). In \cite{lorentzdirac,damping} it was shown that, with this information, the standard boundary condition (\ref{stringbc}) for the string endpoint can be rewritten in the form of an equation of motion for the dressed quark,
\begin{equation}\label{eom}
{d\over d\tau}\left(\frac{m{d x^{\mu}\over d\tau}-{\sqrt{\lambda}\over 2\pi m} \cF^{\mu}}{\sqrt{1-{\lambda\over 4\pi^2 m^4}\cF^2}}\right)=\frac{\cF^{\mu}-{\sqrt{\lambda}\over 2\pi m^2} \cF^2 {d x^{\mu}\over d\tau}}{1-{\lambda\over 4\pi^2 m^4}\cF^2}~.
\end{equation}

Notice that the characteristic length scale appearing in (\ref{eom}) is precisely $z_m=\sqrt{\lambda}/2\pi m$, which as previously discussed, plays the role of the quark Compton wavelength.
Let us now examine the behavior of a quark that is sufficiently heavy, or is forced sufficiently softly, that the condition $\sqrt{\lambda |\cF^2|}/2\pi m^2\ll 1$  holds. It is then natural to expand the equation of motion in a power series in this small parameter. To zeroth order in this expansion, we correctly reproduce the pointlike result
$m \p_{\tau}^2 x^{\mu}=\cF^{\mu}$. If we instead keep terms up to first order, we find
  $$
  m {d\over d\tau}\left( {d x^{\mu}\over d\tau}-{\sqrt{\lambda}\over 2\pi m^2}\cF^{\mu}\right)\simeq\cF^{\mu}-{\sqrt{\lambda}\over 2\pi m^2}\cF^2 {d x^{\mu}\over d\tau}~.
  $$
  In the $\cO(\sqrt{\lambda})$ terms it is consistent, to this order, to replace $\cF^{\mu}$ with its zeroth order value, thereby obtaining
  \begin{equation}\label{ourald}
  m \left( {d^2 x^{\mu}\over d\tau^2}-{\sqrt{\lambda}\over 2\pi m}{d^3 x^{\mu}\over d\tau^3}\right)\simeq\cF^{\mu}-{\sqrt{\lambda}\over 2\pi}{d^2 x^{\nu}\over d\tau^2}{d^2 x_{\nu}\over d\tau^2} {d x^{\mu}\over d\tau}~.
  \end{equation}
  Interestingly, this coincides \emph{exactly} with the Lorentz-Dirac equation (\ref{ald}), with the Compton wavelength (\ref{zm}) playing the role of characteristic size $t_e$ for the composite quark. This is indeed the natural quantum scale of the problem. The radiation reaction force in (\ref{ourald}) is correctly given by the
  covariant Lienard formula, as expected from the  result (\ref{emikh}) \cite{mikhailov}, which we see then arising as
  the pointlike limit of the full radiation rate encoded in the right-hand side of (\ref{eom}).
 The
  Schott term in (\ref{ourald}) (associated with the near field of the quark), originated from the terms inside the $\tau$-derivative in the left-hand side of (\ref{eom}), which we understand then to codify a modified dispersion relation for our composite quark.

To second order in $\sqrt{\lambda |\cF^2|}/2\pi m^2$, we similarly obtain
\begin{eqnarray}\label{ourald2}
m{d^2 x^{\mu}\over d\tau^2}
-{\sqrt{\lambda}\over 2\pi}\left(\overbrace{d^3 x^{\mu}\over d\tau^3}
\underbrace{-{d^2 x^{\nu}\over d\tau^2}{d^2 x_{\nu}\over d\tau^2} {d x^{\mu}\over d\tau}}\right)
&{}&{}\nonumber\\
+{\lambda\over 4\pi^2 m}\left(\overbrace{{d^4 x^{\mu}\over d\tau^4}
-(1}\underbrace{+2){d^2 x^{\nu}\over d\tau^2}{d^3 x_{\nu}\over d\tau^3} {d x^{\mu}\over d\tau}}\right)
&{}&{}\\
-{\lambda\over 4\pi^2 m}\left(\overbrace{1\over 2}+\underbrace{1}\right){d^2 x^{\nu}\over d\tau^2}{d^2 x_{\nu}\over d\tau^2} {d^2 x^{\mu}\over d\tau^2}
&\simeq&\cF^{\mu}~. \nonumber
\end{eqnarray}
For compactness, we have grouped together all terms arising at the same order in the expansion, and used underbraces to mark the radiation reaction terms originating from the right-hand side of (\ref{eom}) (which now include corrections beyond the standard Lienard formula), and overbraces to indicate the near-field terms arising from the left-hand side of (\ref{eom}) (which incorporate corrections to the standard Schott term).

One can continue this expansion procedure to arbitrarily high order in the ratio $\sqrt{\lambda |\cF^2|}/2\pi m^2$.
The full equation (\ref{eom}) is thus recognized as a compact (reduced-order) rewriting of an infinite-derivative extension of the Lorentz-Dirac equation that automatically incorporates the size $z_m$ of our
non-pointlike non-Abelian source. It is curious to note that (\ref{eom}), which clearly incorporates the effect of radiation damping on the quark, has been obtained from (\ref{mikhsol}), which does \emph{not} include such damping for the string itself. The latter arises from the backreaction of the closed string fields set up by our macroscopic string, but these are of order $1/N_c^2$, and therefore
subleading at large $N_c$ (though not necessarily at the real-world value $N_c=3$ \cite{syz}).

The full physical content of (\ref{eom}) can be made transparent by rewriting it in the form
 \begin{equation}\label{eomsplit}
 {d P^{\mu}\over d\tau}\equiv {d p_q^{\mu}\over d\tau}+{d P^{\mu}_{{\mbox{\scriptsize rad}}}\over d\tau}=\cF^{\mu},
 \end{equation}
 recognizing $P^{\mu}$ as the total string ($=$ quark $+$ radiation) four-momentum,
 \begin{equation}\label{pq}
 p_q^{\mu}=\frac{m{d x^{\mu}\over d\tau}-{\sqrt{\lambda}\over 2\pi m} \cF^{\mu}}{\sqrt{1-{\lambda\over 4\pi^2 m^4}\cF^2}}
 \end{equation}
 as the intrinsic momentum of the quark including the contribution of the near-field sourced by it
 (or, in quantum-mechanical language,
  of the gluonic cloud surrounding the quark), and
\begin{equation}\label{radiationrate}
{d P^{\mu}_{{\mbox{\scriptsize rad}}}\over d\tau}={\sqrt{\lambda}\, \cF^2 \over 2\pi m^2}\left(\frac{{d x^{\mu}\over d\tau}-{\sqrt{\lambda}\over 2\pi m^2} \cF^{\mu} }{1-{\lambda\over 4\pi^2 m^4}\cF^2}\right)
\end{equation}
as the rate at which momentum is carried away from the quark by chromo-elec\-tro\-mag\-ne\-tic radiation.

Unlike its classical electrodynamic counterpart  (\ref{ald}), the dressed quark equation of motion (\ref{eom}) has no self-accelerating (runaway) solutions: in the (continuous) absence of an external force, it uniquely predicts that the 4-acceleration of the quark must vanish. Interestingly, the converse to this last statement is not true: constant 4-velocity does not uniquely imply a vanishing force.

It is natural to expect the energy split achieved in (\ref{emikh}) or (\ref{eomsplit}) to be somehow reflected in the geometry of the string worldsheet. It is interesting then that, on the string worldsheet dual to a quark undergoing arbitrary accelerated motion, there appears a black hole \cite{dragtime}, whose event horizon naturally divides the worldsheet into two causally disconnected regions. The appearance of a worldsheet black hole had also been noted previously at finite temperature \cite{gubserqhat,ctqhat}, so the overall lesson is that such causal structure
is intrinsically tied to energy dissipation, be it within a thermal plasma
or in vacuum.  It seems plausible then to interpret the regions outside and inside the black hole as corresponding respectively to the quark and the gluonic field, and postulate that the energy flow across this horizon should be related to the energy radiated by the quark \cite{dominguez,dragtime}. This possibility has been studied more closely in \cite{xiao,beuf}. Unfortunately, this interpretation is not always correct: for general quark trajectories, one can show that the energy of the portions of the string outside and inside the dynamical horizon does not actually match the intrinsic and radiated energy of the quark \cite{eric}.

\subsection{Absence of broadening in the radiation pattern}
\label{broadeningsec}

In the previous two subsections we have seen how a study of the string embedding yields important quantitative information on the dual quark dynamics, including its rate of energy loss, based on the natural split achieved in \cite{mikhailov}
of the total energy of the string. The latter is conserved on the fixed AdS (or AdS-Schwarzschild)
background, but would of course decrease steadily if we take into account the gravitational
(and dilatonic, etc.) radiation given off by the string in the course of its evolution.
As we have noted, from the gravity perspective this radiation is suppressed by a factor of $1/N_c^2$, so it is consistent to neglect it at large $N_c$. On the other hand, through the GKPW recipe for correlation functions \cite{gkpw}, it is this radiation (or,
more precisely, the full metric perturbation produced by the string), evaluated at the AdS
boundary, that determines the expectation value of the gauge theory energy-momentum tensor.
This tensor contains not just the gross information about the total energy loss rate, but also the fine details about the
directionality of the flow and the relative weight of the various dissipation channels.
The problem of determining the spacetime profile of the gauge field disturbance produced by an arbitrary charge trajectory was solved long ago by Lienard and Wiechert for classical electrodynamics, but, under various guises, remains of great interest today in the context of quantum non-Abelian gauge theories. It is therefore significant that the gauge/gravity duality has given us a useful handle on this problem for a varied class of gauge theories in the previously inaccessible regime of strong coupling.

 The translation between string and gauge theory disturbances was first explored in \cite{cg}, which studied the dilaton waves given off by small
fluctuations on an otherwise static, radial string in AdS$_5$, and inferred from them the profile of the dual gluonic field observable
$\expec{\tr F^2(x)}$ in the presence of an oscillating quark in vacuum.  The MSYM waves  were found to display significant temporal broadening, just as one would expect given that points on the non-Abelian field arbitrarily far from the quark can themselves reradiate. This feature emerges naturally in the gravity side of the correspondence, because motion of the string endpoint generates waves that move up along the body of the string, and each point on the string then emits a dilaton wave that travels back down to the observation point $x$ on the AdS boundary, where, via the AdS/CFT recipe for correlation functions \cite{gkpw}, the value of
$\expec{\tr F^2(x)}$ is deduced by assembling together all such contributions, each with a different time delay.
While informative, the results of \cite{cg} did not allow a definite identification of waves with the characteristic $1/|\vec{x}|^2$ falloff associated with
radiation, i.e., contributions that transport energy to infinity. The unambiguous detection of the latter calls for examination of the MSYM energy-momentum tensor, $\expec{T_{\mu\nu}(x)}$, which in the gravity side of the correspondence requires a determination of the gravitational waves emitted by the string \cite{mo}.

In more recent years, motivated by potential contact with the phenomenology of the quark-gluon plasma \cite{qgprev}, analyses of both  $\expec{\tr F^2(x)}$ and $\expec{T_{\mu\nu}(x)}$ have been carried out in the case of a heavy quark moving at constant velocity through a thermal plasma (the only finite-temperature case where an exact solution for the corresponding string embedding is available \cite{hkkky,gubser}), in a large body of work that includes \cite{gluonicprofile} and has been reviewed in \cite{gluonicprofilerev}. The results are again compatible with the expected nonlinear dynamics of the (in this case, finite-temperature) MSYM medium.

Given these antecedents, it came as a surprise when, back at zero temperature, additional calculations  going beyond the linearized string approximation found, first for special cases \cite{liusynchrotron,iancu1} and then for an arbitrary quark trajectory \cite{iancu2} (using the string embedding (\ref{mikhsol})), that the ensuing energy density $\expec{T_{00}(x)}$ displays no temporal broadening, and is in fact as sharply localized in spacetime as the corresponding classical profile.

 The apparent tension between the results of \cite{cg} and \cite{iancu1,iancu2} for the gluonic fields in vacuum was resolved recently in \cite{trfsq}, which carried out the $\expec{\tr F^2(x)}$ calculation beyond the linearized string approximation, and for an arbitrary quark trajectory, to put it on a par with the computation in \cite{iancu2}. The results show that the gluonic near field at any given observation point depends only on dynamical data evaluated at a single retarded time along the quark trajectory (with luminal and subluminal propagation in the cases of infinite and finite quark mass, respectively). This proves that, despite appearances, there is in fact no conflict between \cite{cg} and \cite{iancu1,iancu2}. While it is true that the gluonic field emerges as a superposition of contributions with all possible time delays (as shown explicitly  by the calculations in \cite{liusynchrotron,iancu1,iancu2,trfsq}, and also by a recent reformulation of these as a super
 position of gravitational shock waves emitted by each point along the string \cite{veronika}), it turns out that the \emph{net} result evidences only the smallest of these delays. So, contrary to
 \cite{iancu1,iancu2}, where it was suggested that the supergravity approximation is somehow leaving out the expected non-Abelian rescattering, in \cite{trfsq} the no-broadening result was interpreted  as a surprising \emph{prediction} of the AdS/CFT correspondence for the \emph{net} pattern of propagation in the CFT at strong coupling and with a large number of colors, under the assumption of a purely outgoing condition for the gluonic field generated by the quark.

We should perhaps stress that this prediction refers only to propagation in the CFT vacuum. When a beam of radiation enters a plasma, it is expected to remain unbroadened only at distances that are small compared to the thermal wavelength, after which it will diffuse and eventually thermalize, consistent with the results of \cite{gluonicprofile}.

\subsection{Fluctuations and the Unruh effect}
\label{unruhsec}

When going beyond the classical description of the string, two new effects are found, both of which are suppressed by a factor of the string length divided by the AdS curvature radius, or, equivalently (via (\ref{metric})), by an inverse factor of the CFT coupling $\lambda$. On the one hand, we pick up the usual quantum fluctuations arising from the determinant of the path integral over string embeddings. These are present even for a static string (see, e.g., \cite{fluct}), and lead to spontaneous deviations from the average endpoint/quark trajectory of the type studied, e.g., in \cite{johnson2}. On the other hand,  the worldsheet black hole emits Hawking radiation, which populates the various modes of oscillation of the string.

This second effect, which is present only for an accelerated trajectory and is thus associated with the quantum fluctuations induced by the gluonic radiation emitted by the quark, has been studied in \cite{brownian}, for the special case of uniform proper acceleration $A$.
 In this case, the general solution (\ref{mikhsol}) takes a form that was found independently in \cite{xiao,ppz}. The geometry induced on the worldsheet contains a black hole that is static, and thus amenable to explicit calculation. The analysis of the effect of Hawking radiation is in complete parallel with the story of Brownian motion \cite{rangamani,sonteaney} recalled in Section \ref{latetimesec}, except that it is complicated by the explicit time dependence present in the worldsheet geometry when presented in inertial coordinates.

 The problem simplifies if instead of working in the coordinates appropriate for an inertial observer we transform to a Rindler coordinate system adapted to an observer sitting on the quark. This transformation gives rise to an acceleration horizon both in the boundary and bulk descriptions. As a result, a state that is pure from the inertial perspective will generally be mixed from the point of view of the Rindler observers, because the field degrees of freedom accessible to the latter will be entangled with degrees of freedom in the region beyond their horizon, which they must trace over. In particular, the pure AdS geometry expressed in Rindler coordinates, which is dual to the Minkowski vacuum of the CFT (as evidenced by the vanishing of the expectation value of the stress-energy tensor), is interpreted as a thermal bath at the temperature  $T_{\mathrm{U}}=A/ 2\pi$.
 Contact is thus made with the celebrated Unruh effect \cite{unruh}, i.e., the fact that, from the point of view of a uniformly accelerated observer, the Minkowski vacuum behaves as a thermal medium.
 In Rindler coordinates, the string embedding is static and bends towards the Rindler horizon.
 It is well-known that the Rindler horizon of the CFT can be removed via a Weyl transformation, leading to the open Einstein universe. The corresponding bulk transformation  drastically alters the radial foliation of the AdS geometry. Even though, by construction, in this new conformal frame the acceleration horizon is no longer visible in the boundary description, one can show that it is still present in the bulk, but lies at the fixed radial position that a\-ccor\-ding to the AdS/CFT dictionary corresponds to the Unruh temperature. In other words, after this second transformation, the thermal character of the CFT state arises not from entanglement with degrees of freedom that lie beyond a spacetime horizon, but from the direct identification of the specific energy scale as the temperature of the CFT, in exact parallel with the dual interpretation of the Schwarzschild-AdS geometry. This exercise thus sheds light on the AdS implementation of the Unruh effect. (Related analyse
 s can be found in \cite{xiao,ppz,marianoangel}.)

After the bulk diffeomorphism that removes the CFT horizon, the string  embedding is not only static but also completely vertical. Interestingly, both the base string embedding  and the background geometry  at the location of the string are found to coincide exactly with the $d=2$ \emph{thermal} setup analyzed in \cite{rangamani}, which allows one to obtain the radiation-induced fluctuations of the quark simply by translating the results of that work. This close relation between the quantum fluctuations of the uniformly accelerated quark on Minkowski spacetime and the thermal fluctuations of a static quark in a thermal medium is evidently a direct consequence of the Unruh effect, but the reader should be aware that, for $d>2$, the detailed properties of this thermal medium are found to differ from those of the familiar homogeneous and isotropic thermal ensemble dual to the Schwarzschild-AdS$_{d+1}$ geometry. Back in the original inertial frame, the quark fluctuations are gove
 rned by Langevin type equations \cite{brownian}.

\section{Isolated Quark at Finite Temperature}
\label{qplasmasec}

\subsection{Constant velocity} \label{constantvsec}

Having understood the rate of energy loss for a heavy
quark that moves in the MSYM vacuum, in this section we restore $z_h<\infty$--- and
consequently $h<1$--- in the metric (\ref{metric}), to study the same quantity in the case
where the quark moves through a thermal plasma.
A thorough generalization of Mikhailov's analytic results \cite{mikhailov} to this finite
temperature setup would require finding the exact solution to the Nambu-Goto equation of
motion for the string on the AdS-Schwarzschild background, for any given trajectory of the
string endpoint at $z_m\ge 0$. Sadly, this has not yet been accomplished.
Nevertheless, based on the results discussed in the previous section, we expect the total
energy of the string at any given time to again decompose into a surface term that encodes
the intrinsic energy of the quark and an integrated local term that reflects the energy lost
by the quark to the thermal medium.

There are two easy cases where one can show analytically that this expectation is borne out \cite{dragtime}: the static quark, where there is of course no energy loss, and the quark moving at constant velocity.
An important difference with respect to the $T=0$ case analyzed in
the previous section is that here the surface contribution
arises not only from the
lower ($z=z_m$) but also from the upper ($z=z_h$) endpoint of the
string.
Since $z_h$ marks the position of an event horizon, for any finite coordinate time $t$ the value
of the surface contribution at $z_h$ is not influenced by the behavior of the $z<z_h$
portion of the string, but depends only on the string's configuration at $t\to-\infty$. The
same interpretation can be then carried over to the gauge theory: for arbitrary finite-temperature configurations, the terms in the quark dispersion relation arising from the upper string endpoint will encode a contribution to the energy of the state that depends solely on the initial
configuration of the quark$+$plasma system.

\subsection{Early-time energy loss} \label{acceleratedsec}

 As we mentioned in Section \ref{latetimesec}, expression  (\ref{elossgubser}) was derived under a number of simplifying assumptions. The one that matters most for our purposes is the restriction made by the authors of \cite{hkkky,gubser} to a configuration where the quark is either moving with constant (possibly relativistic) velocity as a result of being pulled by an external force that precisely balances the drag, or is unforced but moving nonrelativistically and about to come to rest. Either of these scenarios requires the interaction between the quark and the medium to occur over a considerable period of time.
The actual energy loss might thus be expected to differ from (\ref{elossgubser}) in a situation where the
  quark moving through the plasma is accelerating, or in the initial
  period following its production within the thermal medium.
  This point has been emphasized from the phenomenological perspective
  in \cite{peigne1}. The estimates there are based on perturbative
  calculations, so it is interesting to inquire into this issue in
  the strongly-coupled systems available to us through
  the AdS/CFT correspondence.

  Now,  the restriction in  \cite{hkkky,gubser}
  to the stationary or asymptotic cases
   was of course implemented
  to gain analytic control on the problem of energy loss.
  The authors of \cite{beuf} were able to slightly extend the stationary result to the case of a slowly decelerating quark, working in an expansion in powers of $\sqrt{\lambda}T/m$.
  Away from these regimes, it is difficult to follow
  the evolution of the quark in
  the thermal plasma, or equivalently, of the string
  on the AdS black hole geometry.

 The first gauge/gravity exploration of early-time energy loss was carried out in \cite{dragtime}, by numerically integrating
the Nambu-Goto equation of motion for the  string dual to a quark
initially at rest, which is accelerated along one dimension by a time-dependent
external force that is turned off after a short interval,
allowing the quark to move thereafter only under the influence of
the plasma.
The results show a qualitative difference between the initial
stage ($0\le t < t_{\mbox{\scriptsize release}}$) where the quark
is accelerated by means of the external force $F(t)$, and the
second stage ($t_{\mbox{\scriptsize release}}\le t
<t_{\mbox{\scriptsize breakdown}}$) where it moves only under the
influence of the plasma.
In the former stage,
the rate of energy loss, for values of $m$ in the neighborhood
of the charm or bottom masses, is in fact nearly identical
to the corresponding rate in vacuum, given by the modified Lienard formula
(\ref{radiationrate}). In the latter
stage, which would appear to be more relevant from the
phenomenological perspective, the quark was found to dissipate
energy at a rate much lower than the late-time result (\ref{elossgubser}). Indeed, for the values $z_m/z_h=0.2,0.3,0.4$ that (as noted below (\ref{zmnoplasma})) are appropriate for the charm and bottom quarks,
  the late-time
friction coefficient $\mu\equiv -(1/p_q)(dp_q/dt)$ is respectively $\mu_{\mbox{\scriptsize
late}}/\pi T=0.25,0.41,0.59$, but the numerical results for $v(t)$ at $t_{\mbox{\scriptsize release}}\le t
<t_{\mbox{\scriptsize breakdown}}$
are best approximated by $\mu_{\mbox{\scriptsize early}}/\pi
T=0.08,0.15,0.26$.
A more detailed discussion of the energy lost by the quark
in both stages was given in \cite{dragtime}, and a similar comparison of vacuum versus thermal energy loss rates was performed later in \cite{liustirring}, for the special case of a quark undergoing uniform circular motion.

Additional control on the problem of early-time quark damping in a strongly-coupled plasma was gained recently in \cite{dampingtemp}, which studied thermal effects analytically, as a small perturbation on the zero-temperature evolution. This is reasonable on general physical grounds during the initial stage of motion through the plasma, and is moreover supported by the numerical results of \cite{dragtime}.
Interestingly, this framework naturally makes contact with the previously known limiting velocity $v_m$ (defined in (\ref{vm})) for a quark immersed in the strongly-coupled plasma.
The results of \cite{dampingtemp} identify
 \begin{equation}\label{eqtemp}
 E_q=\frac{mv_m}{\sqrt{v_m^2-v^2}}-\frac{1}{2}\sqrt{\lambda}T
 \end{equation}
as the thermally corrected intrinsic energy of the quark, and
 \begin{equation}\label{eradtemp}
 {d E_{\mbox{\scriptsize rad}}\over dt}=\frac{\pi^4T^4z_m^3m v_m v}{2(v_m^2-v^2)}
 =\frac{\pi\lambda^{3/2}T^4 v_m v}{16m^2(v_m^2-v^2)}
 \end{equation}
as its rate of energy loss.

For parameter values appropriate to RHIC, this early-time rate can be up to six times \emph{smaller} than the late-time result (\ref{elossgubser}). This comparison is consistent with the (more limited) numerical results of \cite{dragtime} described above. The functional form of the results is also of interest: from the derivation in \cite{dampingtemp} it becomes clear that the $T^4$-dependence seen in (\ref{eradtemp}) (and anticipated in an estimate based on saturation physics \cite{dominguez}) follows naturally from the first thermal corrections in the metric, and via dimensional analysis, this fixes the accompanying dependence on the quark mass, which in turn explains the peculiar power of the 't~Hooft coupling. Another prominent feature of this formula is its velocity-dependence: whereas the late-time friction coefficient deduced from (\ref{elossgubser}) is momentum-independent,
 its early-time counterpart  depends strongly on momentum. This feature should constitute an interesting experimental signature when used as input for phenomenological models, as in \cite{horowitz,kharzeev}.

The rate (\ref{eradtemp}) is limited to the initial stage where the quark is only starting to feel the effects of the plasma, and should be reliable for times that are not too close to the lifetime of the real-world QGP. This suggests that the experimental results should effectively arise from some sort of average or interpolation between the early and late energy loss rates (\ref{eradtemp}) and
 (\ref{elossgubser}).  For times of order $1/\pi T \sqrt{v_m^2-v_0^2}$, one expects the string embedding to be significantly modified by the appearance of $T^8$ and higher corrections neglected in the linearized analysis of \cite{dampingtemp}. As the string continues to evolve, a dynamical worldsheet horizon (located at $z=z_h$ in the remote past) is expected to move up toward smaller values of $z$, and then (if the quark is undisturbed) move back down towards the spacetime horizon \cite{dragtime}. When the wavefront on the string crosses the worldsheet horizon, one would have to deviate from the purely outgoing condition used in \cite{dampingtemp}, to accommodate the fact that, as the medium is disturbed by the quark, it is expected to radiate gluonic waves back towards the source. The net result should be to yield a damping coefficient $\mu(t)$ that smoothly interpolates between the early-time result and the late-time form.

\section{Pair Creation within the Plasma}
\label{qqbarsec}

\subsection{Quark-antiquark evolution}

As was mentioned in Section \ref{latetimesec}, in the experimental setup the heavy quark is created within the plasma together with its corresponding antiquark, and the presence of the latter would be expected to substantially modify the gluonic fields in the vicinity of the quark, consequently affecting its evolution. The evolution of
a heavy quark and antiquark that are created within the plasma and
then separate back to back was first studied in \cite{hkkky}, and explored in greater depth in \cite{dragtime}. In dual language, this corresponds to a string with both of its endpoints on the D7-branes at $z=z_m$, such that the endpoints are initially at the same spatial location but have initial velocities in opposite directions.

In setting up the problem, one realizes that on the gauge theory side there are actually two distinct quark-antiquark configurations:  the product of a fundamental $q$ and an antifundamental $\bar{q}$ can lead
to a $q$-$\bar{q}$ pair either in the singlet or the adjoint representation of the $SU(N_c)$
gauge group. There is a counterpart to this in the gravity side, because there are precisely two distinct types of consistent initial conditions for the string, which lead it to expand into either a $\cup$-shape or a $\vee$-shape as its endpoints separate \cite{dragtime}.
Curiously, the initial quark velocity is found to be freely
adjustable in the adjoint, but not the singlet, configuration--- in the latter case it is invariably fixed at the limiting value $v=v_m$.

\begin{figure}[htb]
\begin{center}
\vspace*{0.2cm} \setlength{\unitlength}{1cm}
\includegraphics[width=9cm,height=6cm]{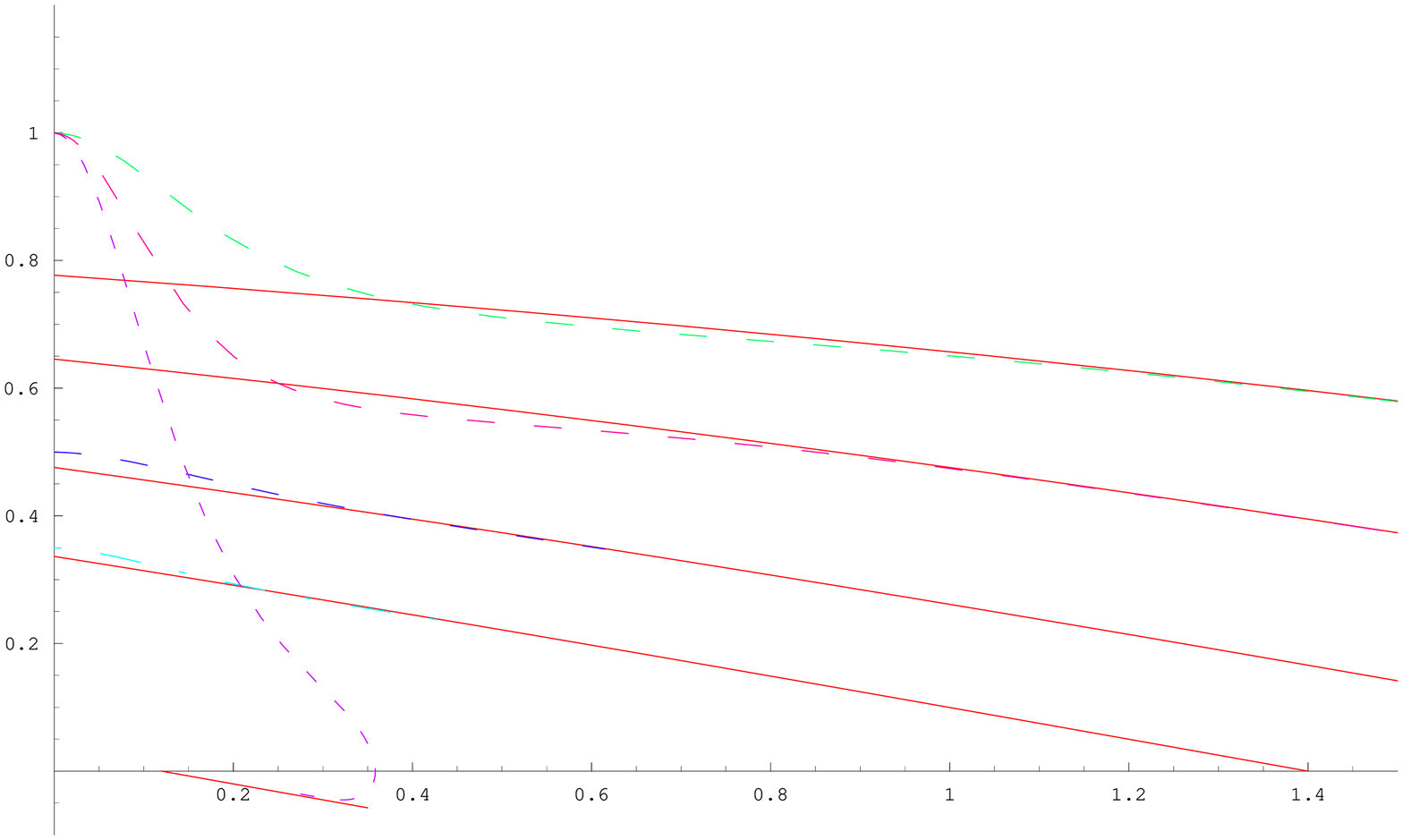}
 \begin{picture}(0,0)
   \put(0.1,0.4){$x$}
   \put(-8.7,6.1){$v$}
   \put(-6.8,5){\vector(0,-1){1}}
    \put(-4.9,4.0){\vector(0,-1){1}}
   \put(-6.8,1.4){\vector(0,-1){1}}
  \put(-8.7,3.8){\vector(0,-1){1}}
   \put(-8.7,1.0){\vector(0,1){1}}
    \end{picture}\hspace{1cm}
\caption{Quark evolution (velocity as a function of traveled distance,
in units of $1/\pi T$)
 for five different initial conditions. The dotted curves show the results
 of the numerical integration in \cite{dragtime}, contrasted against fits in solid red that use
  the late-time analytic expression (\ref{elossgubser}). The three dotted curves starting
  at $v_0=v_m$ describe singlet configurations with different
  initial velocity profiles and energies.
Notice that the purple curve describes a
situation where the quark and antiquark turn around and come to rest while approaching
one another. The two remaining curves arise from adjoint configurations with different energies and initial quark
velocities. The vertical arrows mark the transition distance where each trajectory enters the late-time regime.} \label{stage2fig}
\end{center}
\end{figure}

As the quark and antiquark separate, we expect them to eventually enter the late-time regime where their rate of energy loss is given by (\ref{elossgubser}).
The agreement between the analytic and late-time numeric results is
most cleanly seen if instead of comparing graphs of $x(t)$ or
$v(t)$ for the quark (where one would need to look at
$t\to\infty$), one examines the plots of $v(x)$. These plots are shown in Fig.~\ref{stage2fig}, for
  mass parameter $z_m/z_h=0.2$ (in the neighborhood
 of the charm quark), corresponding
  to a limiting velocity $v_m=0.9992$. It is
evident from the figure that the late-time behavior is well-described by
(\ref{elossgubser}) in all cases, but there is also an initial period
where the behavior is different.
 This difference is clearly more significant for
 the singlet than the adjoint case, just as one would expect. For values of the mass in the neighborhood of the charm or bottom masses, the numerical results of \cite{dragtime} indicate that the initial evolution is essentially identical to what it would be in the absence of the plasma.
  This picture seems rather close to the phenomenological
  discussion given
  in \cite{peigne1} (in the context of collisional
 energy loss):
  when the singlet quark-antiquark pair is formed
 within the plasma, there is a delay before
 the interaction between the newly created sources and the plasma
 can set up the long range gluonic field profile that is responsible
 for the late-time dissipation.
To examine in more detail the transition to the late-time behavior,  for a variety of trajectories one can
determine the point $(x_f,v_f)$ beyond which the numeric $v(x)$
curve agrees with the late-time curve to a given
accuracy $f$. (The work \cite{dragtime} considered $f=5$ or 10\%.)
These transition points are schematically
indicated by the arrows in Fig.~\ref{stage2fig}.

\subsection{Screening length}

 By the time when the quark moves beyond $x_f(v_f)$ and therefore
 enters the late-time regime, it is certainly insensitive to the presence of the antiquark. It is natural then to ask how the transition distance $x_f(v_f)$ compares against (half of) the length $L_{\mbox{\scriptsize max}}(v)$ beyond which the quark and antiquark are screened from each other by the plasma. Based on what we just said, we know that for a given velocity we must have $L_{\mbox{\scriptsize max}}/2 <x_f$, but the actual comparison informs us on whether the transition to the regime
 where the quark experiences a constant drag coefficient occurs
 right after the quark and antiquark are screened from each other
 by the plasma, or if there is an intermediate regime where the
 quark moves independently from the antiquark but nevertheless
 feels a drag force that differs from the stationary result of \cite{hkkky,gubser,ct},
 as we found when studying the early-time evolution of an isolated quark in Section
 \ref{acceleratedsec}.

The screening length for infinitely massive quarks in MSYM was computed in \cite{liuwind,dragqqbar}, by considering a quark-antiquark pair moving jointly through the plasma (a related length was plotted in
 \cite{sonnenschein}, and \cite{dragqqbar} gave in addition an extensive analysis of the energy of the $q$-$\bar{q}$ pair). This entails a study of a $\cup$-shaped string whose endpoints move in the same direction along the Schwarzschild-AdS geometry, a configuration that, interestingly, feels no drag in the $N_c\to\infty$ limit, on account of its being color-neutral \cite{sonnenschein,liuwind,dragqqbar}.\footnote{The $\cO(1/N_c^2)$ drag force has been worked out by indirect means in \cite{dusling}.} Over
the entire range $0\le v\le 1$ the screening length may be approximated
as \cite{dragqqbar}
\begin{equation} \label{onethird}
L_{\mbox{\scriptsize max}}(v)\approx{0.865\over \pi
T}(1-v^2)^{1/3}~,
\end{equation}
while in the ultra-relativistic limit, it can be shown analytically that \cite{liuwind}
\begin{equation}\label{onequarteranal}
L_{\mbox{\scriptsize max}}(v)\to {1\over \pi T}{
3^{-3/4}4\pi^{3/2}\over\Gamma(1/4)^2}(1-v^2)^{1/4}\simeq
{0.743\over \pi T}(1-v^2)^{1/4}\quad\mbox{for $v\to 1$~.}
\end{equation}
The $1/4$ in the exponent here has an intuitive interpretation in terms of the boosted energy density of the thermal medium \cite{liuwilson}.
The full curve $L_{\mbox{\scriptsize max}}(v)$ does not deviate
far from this asymptotic form, so a decent
approximation to it is obtained by replacing $0.743\to 0.865$ in
(\ref{onequarter}), to reproduce the correct value at $v=0$ (at
the expense of introducing a 16\% error as $v\to 1$) \cite{liuwind}:
\begin{equation} \label{onequarter}
L_{\mbox{\scriptsize max}}(v)\approx{0.865\over \pi
T}(1-v^2)^{1/4}~.
\end{equation}

\begin{figure}[htb]
\vspace*{0.5cm}
\begin{center}
\setlength{\unitlength}{1cm}
\includegraphics[width=6cm,height=4cm]{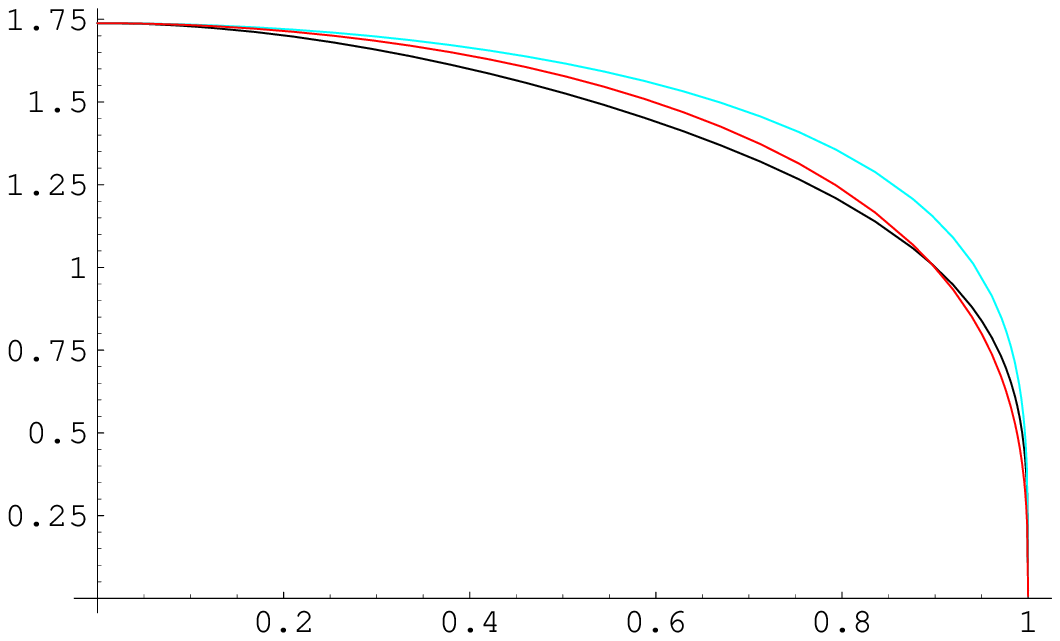}
 \begin{picture}(0,0)
   \put(0.2,0.3){$v$}
   \put(-6.1 ,4.2){$L_{\mbox{\scriptsize max}}$}
 \end{picture}
\caption{Screening length $L_{\mbox{\scriptsize max}}$ (in units
of $1/2\pi T$) as a
function of velocity (in black) compared against the
approximations (\ref{onethird}) (in red) and (\ref{onequarter}) (in
blue).}\label{onethirdfig}
\end{center}
\end{figure}

A comparison between the two approximations (\ref{onethird})
 and (\ref{onequarter}) is shown
 in Fig.~\ref{onethirdfig}: overall, the
 exponent $1/3$
is better than $1/4$ in the sense that
it implies a smaller squared deviation from the numerical results,
even though $1/4$ leads
to a smaller \emph{percentage} error in the range $v>0.991$
($\gamma>7.3$).
An attempt to better parametrize the deviation away from the
ultra-relativistic behavior was made in \cite{sfetsosqqbar}.
In any case, one should bear in mind that the region of principal
interest at RHIC is not really $v\to 1$, but
 $\gamma v\sim 1$.

To compare against the transition distance $x_f$ defined in the previous subsection for the back-to-back quark and antiquark, the authors of \cite{dragtime} extended this calculation to the case of finite mass. The main difference with respect to the $m\to\infty$ case is
 that now the `ultra-relativistic' region would refer to the limit where the pair velocity
 approaches the limiting velocity $v_m<1$.
The screening length
$L_{\mbox{\scriptsize max}}(v)$ is still relatively well
approximated in the full range $0\le v\le v_m$ by the natural modification
of
the $z_m=0$
fit (\ref{onethird}) (and in all cases, the fit analogous
to (\ref{onequarter})
 does a poorer job).
The behavior in the $v\to v_m$ region
 can still be determined analytically, and turns out to be linear in $v_m^2-v^2$.
 Over the full velocity interval, the screening length $L_{\mbox{\scriptsize max}}(v)$ and
the transition distance $x_f(v)$ and (half of) are found to be of comparable magnitude and scale with velocity in a similar manner \cite{dragtime}. A similar agreement was reported in \cite{iancujet}, and later related work may be found in \cite{liustirring}. Notice that the agreement holds in spite of the fact that the two relevant string
configurations are quite different, with motion respectively along and perpendicular to the
plane in which the string extends.

The conclusion then is that the transition to the
late-time regime (\ref{elossgubser}) takes place immediately after the quark and antiquark lose
contact with one another. That is to say, unlike what we found for the forced isolated quark
in Section \ref{acceleratedsec}, here there is no appreciable intermediate stage where the quark and
antiquark decelerate independently from one another at a rate that differs substantially
from the late-time result of \cite{hkkky,gubser}.
This then extends and at the same time delimits the region
where the analytic results of \cite{hkkky,gubser,ct} can justifiably be used to model energy loss in heavy ion collisions.

 The way in which the findings reviewed in the present section interface with those of Section \ref{acceleratedsec} was discussed in \cite{dampingtemp}. The upshot is that, in the more realistic situation where the presence of the accompanying antiquark is taken into consideration, the early-time damping results would be most relevant in the stage before the separating quark and antiquark lose contact with one another. This is natural, because screening emerges only when thermal effects are substantial. For a quark ploughing through a significant portion of the plasma, then, the late-time formula (\ref{elossgubser}) actually stands a chance of controlling the larger portion of the evolution.

\section*{Acknowledgements}
The present work was partially supported by Mexico's National Council of Science and Technology (CONACyT) grant 104649, by DGAPA-UNAM under grant IN110312, and by the National Science Foundation under Grant No. PHY-0969020.
The research of M.Ch. is supported by 2009-SGR-168, MEC FPA2010-20807-C02-01, MEC FPA2010-20807-C02-02, CPAN CSD2007-00042 Consolider-Ingenio 2010. J.F.P. would like to thank the Texas Cosmology Center for partial support.


\begin{thebibliography}{99}

\bibitem{qgprev}
  M.~J.~Tannenbaum,
  ``Results from RHIC with Implications for LHC,''
  arXiv:1006.5701 [nucl-ex];\\
  W.~A.~Zajc,
  ``The Fluid Nature of Quark-Gluon Plasma,''
   Nucl.\ Phys.\  A {\bf 805}, 283 (2008)
  [arXiv:0802.3552 [nucl-ex]];\\
  B.~Muller,
  ``From Quark-Gluon Plasma to the Perfect Liquid,''
  Acta Phys.\ Polon.\  B {\bf 38}, 3705 (2007)
  [arXiv:0710.3366 [nucl-th]];\\
E.~Shuryak,
``Emerging theory of strongly coupled quark-gluon plasma,''
arXiv:hep-ph/0703208.

\bibitem{energylossrev}
  A.~Majumder and M.~Van Leeuwen,
  ``The theory and phenomenology of perturbative QCD based jet quenching,''
  arXiv:1002.2206 [hep-ph];\\
  U.~A.~Wiedemann,
  ``Jet Quenching in Heavy Ion Collisions,''
  arXiv:0908.2306 [hep-ph];\\
  R.~Rapp and H.~van Hees,
  ``Heavy Quark Diffusion as a Probe of the Quark-Gluon Plasma,''
  arXiv:0803.0901 [hep-ph];\\
  J.~Casalderrey-Solana and C.~A.~Salgado,
  ``Introductory lectures on jet quenching in heavy ion collisions,''
  Acta Phys.\ Polon.\  B {\bf 38}, 3731 (2007)
  [arXiv:0712.3443 [hep-ph]].

 \bibitem{malda}
  J.~M.~Maldacena,
  ``The large $N$ limit of superconformal field theories and
  supergravity,''
  Adv.\ Theor.\ Math.\ Phys.\  {\bf 2}, 231 (1998)
  [Int.\ J.\ Theor.\ Phys.\  {\bf 38}, 1113 (1999)]
  [arXiv:hep-th/9711200].

  \bibitem{magoo}
  O.~Aharony, S.~S.~Gubser, J.~M.~Maldacena, H.~Ooguri and Y.~Oz,
   ``Large $N$ field theories, string theory and gravity,''
  Phys.\ Rept.\  {\bf 323}, 183 (2000)
  [arXiv:hep-th/9905111].

\bibitem{hkkky}
  C.~P.~Herzog, A.~Karch, P.~Kovtun, C.~Kozcaz and L.~G.~Yaffe,
  ``Energy loss of a heavy quark moving through $\cN = 4$ supersymmetric
  Yang-Mills plasma,''
  JHEP {\bf 0607} (2006) 013
  [arXiv:hep-th/0605158].

\bibitem{gubser}
  S.~S.~Gubser,
  ``Drag force in AdS/CFT,''
  Phys.\ Rev.\  D {\bf 74} (2006) 126005
  [arXiv:hep-th/0605182].

\bibitem{ct}
  J.~Casalderrey-Solana and D.~Teaney,
  ``Heavy quark diffusion in strongly coupled $\cN = 4$ Yang Mills,''
  Phys.\ Rev.\  D {\bf 74} (2006) 085012
  [arXiv:hep-ph/0605199].

\bibitem{liuqhat}
  H.~Liu, K.~Rajagopal and U.~A.~Wiedemann,
  ``Calculating the jet quenching parameter from AdS/CFT,''
  Phys.\ Rev.\ Lett.\  {\bf 97} (2006) 182301
  [arXiv:hep-ph/0605178].

\bibitem{clmrw}
  J.~Casalderrey-Solana, H.~Liu, D.~Mateos, K.~Rajagopal and U.~A.~Wiedemann,
  ``Gauge/String Duality, Hot QCD and Heavy Ion Collisions,''
  arXiv:1101.0618 [hep-th];\\
  V.~E.~Hubeny and M.~Rangamani,
  ``A holographic view on physics out of equilibrium,''
  Adv.\ High Energy Phys.\  {\bf 2010} (2010) 297916
  [arXiv:1006.3675 [hep-th]];\\
  A.~Karch,
  ``Recent Progress in Applying Gauge/Gravity Duality to Quark-Gluon Plasma Physics,''
  [arXiv:1108.4014 [hep-ph]];\\
  J.~D.~Edelstein, J.~P.~Shock and D.~Zoakos,
  ``The AdS/CFT Correspondence and Non-perturbative QCD,''
  AIP Conf.\ Proc.\  {\bf 1116}, 265 (2009)
  [arXiv:0901.2534 [hep-ph]];\\
  S.~S.~Gubser and A.~Karch,
  ``From gauge-string duality to strong interactions: a Pedestrian's Guide,''
  Ann.\ Rev.\ Nucl.\ Part.\ Sci.\  {\bf 59} (2009) 145
  [arXiv:0901.0935 [hep-th]];\\
  D.~Mateos,
  ``String Theory and Quantum Chromodynamics,''
  Class.\ Quant.\ Grav.\  {\bf 24} (2007) S713
  [arXiv:0709.1523 [hep-th]];\\
  K.~Peeters and M.~Zamaklar,
  ``The string/gauge theory correspondence in QCD,''
  Eur.\ Phys.\ J.\ ST {\bf 152} (2007) 113
  [arXiv:0708.1502 [hep-ph]].

\bibitem{eta}
  G.~Policastro, D.~T.~Son and A.~O.~Starinets,
  ``The shear viscosity of strongly coupled $\mathcal{N} = 4$ supersymmetric Yang-Mills
  plasma,''
  Phys.\ Rev.\ Lett.\  {\bf 87} (2001) 081601
  [arXiv:hep-th/0104066];\\
  A.~Buchel and J.~T.~Liu,
  ``Universality of the shear viscosity in supergravity,''
  Phys.\ Rev.\ Lett.\  {\bf 93}, 090602 (2004)
  [arXiv:hep-th/0311175];\\
  P.~Kovtun, D.~T.~Son and A.~O.~Starinets,
  ``Viscosity in strongly interacting quantum field theories from black hole
  physics,''
  Phys.\ Rev.\ Lett.\  {\bf 94} (2005) 111601
  [arXiv:hep-th/0405231];\\
  A.~Buchel, J.~T.~Liu and A.~O.~Starinets,
  ``Coupling constant dependence of the shear viscosity in $\cN=4$ supersymmetric
  Yang-Mills theory,''
  Nucl.\ Phys.\  B {\bf 707}, 56 (2005)
  [arXiv:hep-th/0406264].


\bibitem{etarev}
  A.~O.~Starinets,
  ``Transport coefficients of strongly coupled gauge theories: Insights from
  string theory,''
  Eur.\ Phys.\ J.\  A {\bf 29} (2006) 77
  [arXiv:nucl-th/0511073];\\
  D.~T.~Son and A.~O.~Starinets,
  ``Viscosity, Black Holes, and Quantum Field Theory,''
  Ann.\ Rev.\ Nucl.\ Part.\ Sci.\  {\bf 57} (2007) 95
  [arXiv:0704.0240 [hep-th]];\\
  R.~B.~Peschanski,
  ``Quark-Gluon Plasma/Black Hole duality from Gauge/Gravity Correspondence,''
  arXiv:0710.0756 [hep-ph].

\bibitem{liu}
  H.~Liu, K.~Rajagopal and U.~A.~Wiedemann,
  ``Calculating the jet quenching parameter from AdS/CFT,''
  Phys.\ Rev.\ Lett.\  {\bf 97} (2006) 182301
  [arXiv:hep-ph/0605178].

\bibitem{sonnenschein}
  K.~Peeters, J.~Sonnenschein and M.~Zamaklar,
  ``Holographic melting and related properties of mesons in a quark gluon
  plasma,''
  Phys.\ Rev.\  D {\bf 74} (2006) 106008
  [arXiv:hep-th/0606195].

\bibitem{liuwind}
  H.~Liu, K.~Rajagopal and U.~A.~Wiedemann,
  ``An AdS/CFT calculation of screening in a hot wind,''
  Phys.\ Rev.\ Lett.\  {\bf 98} (2007) 182301
  [arXiv:hep-ph/0607062].

\bibitem{dragqqbar}
  M.~Chernicoff, J.~A.~Garc\'\i a and A.~G\"uijosa,
  ``The energy of a moving quark-antiquark pair in an $\cN = 4$ SYM plasma,''
  JHEP {\bf 0609} (2006) 068
  [arXiv:hep-th/0607089].

 \bibitem{dusling}
  K.~Dusling, J.~Erdmenger, M.~Kaminski, F.~Rust, D.~Teaney and C.~Young,
  ``Quarkonium transport in thermal AdS/CFT,''
  JHEP {\bf 0810} (2008) 098
  [arXiv:0808.0957 [hep-th]].

\bibitem{draggluon}
  M.~Chernicoff, A.~G\"uijosa,
  ``Energy Loss of Gluons, Baryons and k-Quarks in an N=4 SYM Plasma,''
  JHEP {\bf 0702 } (2007)  084.
  [hep-th/0611155].

\bibitem{liu5}
  C.~Athanasiou, H.~Liu and K.~Rajagopal,
  ``Velocity Dependence of Baryon Screening in a Hot Strongly Coupled Plasma,''
  JHEP {\bf 0805 } (2008)  083
  [arXiv:0801.1117 [hep-th]].

\bibitem{krishnan}
  C.~Krishnan,
  ``Baryon Dissociation in a Strongly Coupled Plasma,''
  JHEP {\bf 0812}, 019 (2008)
  [arXiv:0809.5143 [hep-th]].

  \bibitem{gubsergluon}
  S.~S.~Gubser, D.~R.~Gulotta, S.~S.~Pufu and F.~D.~Rocha,
  ``Gluon energy loss in the gauge-string duality,''
   JHEP {\bf 0810 } (2008)  052
  [arXiv:0803.1470 [hep-th]].

\bibitem{jankarch}
  S.~Janiszewski, A.~Karch,
  ``Moving Defects in AdS/CFT,''
  [arXiv:1106.4010 [hep-th]].

\bibitem{bigazzi}
  F.~Bigazzi, A.~L.~Cotrone, J.~Mas, A.~Paredes, A.~V.~Ramallo and J.~Tarrio,
  ``D3-D7 Quark-Gluon Plasmas,''
  JHEP {\bf 0911} (2009) 117
  [arXiv:0909.2865 [hep-th]].

\bibitem{otherdragforce}
  C.~Hoyos-Badajoz,
  ``Drag and jet quenching of heavy quarks in a strongly coupled $\mathcal{N}=2^*$ plasma,''
  JHEP {\bf 0909} (2009) 068
  [arXiv:0907.5036 [hep-th]];\\
  U.~Gursoy, E.~Kiritsis, G.~Michalogiorgakis and F.~Nitti,
  ``Thermal Transport and Drag Force in Improved Holographic QCD,''
  JHEP {\bf 0912} (2009) 056
  [arXiv:0906.1890 [hep-ph]];\\
  M.~Mia, K.~Dasgupta, C.~Gale and S.~Jeon,
  ``Five Easy Pieces: The Dynamics of Quarks in Strongly Coupled Plasmas,''
  Nucl.\ Phys.\  B {\bf 839} (2010) 187
  [arXiv:0902.1540 [hep-th]];\\
  W.~A.~Horowitz and Y.~V.~Kovchegov,
  ``Shock Treatment: Heavy Quark Drag in a Novel AdS Geometry,''
  Phys.\ Lett.\  B {\bf 680} (2009) 56
  [arXiv:0904.2536 [hep-th]];\\
  A.~Stoffers, I.~Zahed,
  ``Holographic Jets in an Expanding Plasma,''
  [arXiv:1110.2943 [hep-th]];\\
  M.~Ali-Akbari and U.~Gursoy,
  ``Rotating strings and energy loss in non-conformal holography,''
  arXiv:1110.5881 [hep-th].

\bibitem{panero} 
  M.~Panero,
  ``Thermodynamics of the QCD plasma and the large-N limit,''
  Phys.\ Rev.\ Lett.\ \ {\bf 103}, 232001  (2009)
  arXiv:0907.3719 [hep-lat].



\bibitem{horowitz}
  W.~A.~Horowitz,
  ``pQCD vs. AdS/CFT Tested by Heavy Quark Energy Loss,''
  J.\ Phys.\ G {\bf 35} (2008) 044025
  [arXiv:0710.0703 [nucl-th]];\\
  W.~A.~Horowitz,
  ``Testing AdS/CFT at LHC,''
  arXiv:0905.0504 [hep-ph];\\
   W.~A.~Horowitz and M.~Gyulassy,
  ``Quenching and Tomography from RHIC to LHC,''
  arXiv:1107.2136 [hep-ph];\\
  ``Testing AdS/CFT Drag and pQCD Heavy Quark Energy Loss,''
  J.\ Phys.\ G {\bf 35} (2008) 104152
  [arXiv:0804.4330 [hep-ph]];\\
  C.~Marquet and T.~Renk,
  ``Jet quenching in the strongly-interacting quark-gluon plasma,''
  Phys.\ Lett.\  B {\bf 685} (2010) 270
  [arXiv:0908.0880 [hep-ph]].

\bibitem{bngt}
  B.~Betz, J.~Noronha, M.~Gyulassy and G.~Torrieri,
  ``Jet Energy Loss and Mach Cones in pQCD vs. AdS/CFT,''
  PoS  {\bf CONFINEMENT8} (2008) 130
  [arXiv:0812.1905 [hep-ph]];\\
  ``Anomalous Conical Di-jet Correlations in pQCD vs AdS/CFT,''
  Phys.\ Lett.\  B {\bf 675} (2009) 340
  [arXiv:0807.4526 [hep-ph]].

\bibitem{cm}
  J.~Casalderrey-Solana and D.~Mateos,
  ``Prediction of a Photon Peak in Relativistic Heavy Ion Collisions,''
  Phys.\ Rev.\ Lett.\  {\bf 102} (2009) 192302
  [arXiv:0806.4172 [hep-ph]].

  \bibitem{cdm}
  J.~Casalderrey-Solana, D.~Fernandez and D.~Mateos,
  ``A New Mechanism of Quark Energy Loss,''
  Phys.\ Rev.\ Lett.\  {\bf 104} (2010) 172301
  [arXiv:0912.3717 [hep-ph]];\\
  ``Cherenkov mesons as in-medium quark energy loss,''
  JHEP {\bf 1011} (2010) 091
  [arXiv:1009.5937 [hep-th]].

\bibitem{kharzeev}
  D.~E.~Kharzeev,
  ``Universal upper bound on the energy of a parton escaping from the strongly coupled quark-gluon matter,''
  [arXiv:0806.0358 [hep-ph]].

\bibitem{kk}
  A.~Karch and E.~Katz,
  ``Adding flavor to AdS/CFT,''
  JHEP {\bf 0206} (2002) 043
  [arXiv:hep-th/0205236].

\bibitem{unquenched}
   I.~Kirsch, D.~Vaman,
  ``The D3/D7 background and flavor dependence of Regge trajectories,''
  Phys.\ Rev.\  {\bf D72 } (2005)  026007.
  [hep-th/0505164].

\bibitem{gubsercompare}
  S.~S.~Gubser,
  ``Comparing the drag force on heavy quarks in $\cN = 4$ super-Yang-Mills theory
  and QCD,''
  Phys.\ Rev.\  D {\bf 76} (2007) 126003
  [arXiv:hep-th/0611272].

\bibitem{gkpw}
  S.~S.~Gubser, I.~R.~Klebanov and A.~M.~Polyakov,
  ``Gauge theory correlators from non-critical string theory,''
  Phys.\ Lett.\ B {\bf 428}, 105 (1998)
  [arXiv:hep-th/9802109];\\
  E.~Witten,
  ``Anti-de Sitter space and holography,''
  Adv.\ Theor.\ Math.\ Phys.\  {\bf 2}, 253 (1998)
  [arXiv:hep-th/9802150].

\bibitem{martinfsq}
  J.~L.~Hovdebo, M.~Kruczenski, D.~Mateos, R.~C.~Myers and D.~J.~Winters,
  ``Holographic mesons: Adding flavor to the AdS/CFT duality,''
  Int.\ J.\ Mod.\ Phys.\  A {\bf 20} (2005) 3428.

 \bibitem{cg}
  C.~G.~Callan and A.~G\"uijosa,
  ``Undulating strings and gauge theory waves,''
  Nucl.\ Phys.\ B {\bf 565}, 157 (2000)
  [arXiv:hep-th/9906153].

  \bibitem{linshuryak}
  S.~Lin, E.~Shuryak,
  ``Stress tensor of static dipoles in strongly coupled N = 4 gauge theory,''
  Phys.\ Rev.\  {\bf D76}, 085014 (2007)
  [arXiv:0707.3135 [hep-th]].

\bibitem{lightpartonenergyloss}
   S.~S.~Gubser, D.~R.~Gulotta, S.~S.~Pufu, F.~D.~Rocha,
  ``Gluon energy loss in the gauge-string duality,''
  JHEP {\bf 0810 } (2008)  052 
  [arXiv:0803.1470 [hep-th]];\\
   P.~M.~Chesler, K.~Jensen, A.~Karch,
  ``Jets in strongly-coupled $\mathcal{N} = 4$ super Yang-Mills theory,''
  Phys.\ Rev.\  {\bf D79 } (2009)  025021.
  [arXiv:0804.3110 [hep-th]];\\
 P.~M.~Chesler, K.~Jensen, A.~Karch, L.~G.~Yaffe,
  ``Light quark energy loss in strongly-coupled $\mathcal{N} = 4$  supersymmetric Yang-Mills plasma,''
  Phys.\ Rev.\  {\bf D79 } (2009)  125015 
  [arXiv:0810.1985 [hep-th]];\\
 P.~Arnold, D.~Vaman,
  ``Jet quenching in hot strongly coupled gauge theories revisited: 3-point correlators with gauge-gravity duality,''
  JHEP {\bf 1010 } (2010)  099
  [arXiv:1008.4023 [hep-th]];\\
    ``Jet quenching in hot strongly coupled gauge theories simplified,''
  JHEP {\bf 1104 } (2011)  027
  [arXiv:1101.2689 [hep-th]].

  \bibitem{iancujet}
  Y.~Hatta, E.~Iancu and A.~H.~Mueller,
  ``Jet evolution in the $\mathcal{N}=4$ SYM plasma at strong coupling,''
  JHEP {\bf 0805 } (2008)  037
  [arXiv:0803.2481 [hep-th]].

\bibitem{gluonicprofile}
  J.~J.~Friess, S.~S.~Gubser and G.~Michalogiorgakis,
  ``Dissipation from a heavy quark moving through $\cN = 4$ super-Yang-Mills
  plasma,''
  JHEP {\bf 0609} (2006) 072
  [arXiv:hep-th/0605292];\\
  J.~J.~Friess, S.~S.~Gubser, G.~Michalogiorgakis and S.~S.~Pufu,
  ``The stress tensor of a quark moving through $\cN = 4$ thermal plasma,''
  Phys.\ Rev.\  D {\bf 75} (2007) 106003
  [arXiv:hep-th/0607022];\\
  A.~Yarom,
  ``The high momentum behavior of a quark wake,''
  Phys.\ Rev.\  D {\bf 75} (2007) 125010
  [arXiv:hep-th/0702164];\\
  P.~M.~Chesler and L.~G.~Yaffe,
  ``The wake of a quark moving through a strongly-coupled $\mathcal N=4$
  supersymmetric Yang-Mills plasma,''
  Phys.\ Rev.\ Lett.\  {\bf 99} (2007) 152001
  [arXiv:0706.0368 [hep-th]].
  J.~Noronha, G.~Torrieri, M.~Gyulassy,
  ``Near Zone Navier-Stokes Analysis of Heavy Quark Jet Quenching in an $\mathcal{N} = 4$  SYM Plasma,''
  Phys.\ Rev.\  {\bf C78 } (2008)  024903.
  [arXiv:0712.1053 [hep-ph]];\\
  J.~Hong, D.~Teaney, P.~M.~Chesler,
  ``The Wake of a Heavy Quark in Non-Abelian Plasmas : Comparing Kinetic Theory and the AdS/CFT Correspondence,''
  [arXiv:1110.5292 [nucl-th]].

  \bibitem{gluonicprofilerev}
  S.~S.~Gubser, S.~S.~Pufu, F.~D.~Rocha and A.~Yarom,
  ``Energy loss in a strongly coupled thermal medium and the gauge-string
  duality,''
  arXiv:0902.4041 [hep-th].

\bibitem{liusynchrotron}
  C.~Athanasiou, P.~M.~Chesler, H.~Liu, D.~Nickel and K.~Rajagopal,
  ``Synchrotron radiation in strongly coupled conformal field theories,''
  Phys.\ Rev.\  D {\bf 81} (2010) 126001
  [arXiv:1001.3880 [hep-th]].

\bibitem{iancu1}
  Y.~Hatta, E.~Iancu, A.~H.~Mueller, D.~N.~Triantafyllopoulos,
  ``Aspects of the UV/IR correspondence : energy broadening and string fluctuations,''
  JHEP {\bf 1102 } (2011)  065.
  [arXiv:1011.3763 [hep-th]].

 \bibitem{veronika}
  V.~E.~Hubeny,
  ``Relativistic Beaming in AdS/CFT,''
  arXiv:1011.1270 [hep-th];\\
  ``Holographic dual of collimated radiation,''
  New J.\ Phys.\  {\bf 13}, 035006 (2011)
  [arXiv:1012.3561 [hep-th]].

  \bibitem{iancu2}
  Y.~Hatta, E.~Iancu, A.~H.~Mueller, D.~N.~Triantafyllopoulos,
  ``Radiation by a heavy quark in $\mathcal{N} = 4$  SYM at strong coupling,''
  [arXiv:1102.0232 [hep-th]].

  \bibitem{trfsq}
  M.~Chernicoff, A.~G\"uijosa, J.~F.~Pedraza,
``The Gluonic Field of a Heavy Quark in Conformal Field Theories at Strong Coupling,''
  [arXiv:1106.4059 [hep-th]].

  \bibitem{baier}
  R.~Baier,
``On radiation by a heavy quark in $\cN = 4$ SYM,''
  arXiv:1107.4250 [hep-th].

\bibitem{liuwilson}
  H.~Liu, K.~Rajagopal and U.~A.~Wiedemann,
  ``Wilson loops in heavy ion collisions and their calculation in AdS/CFT,''
  JHEP {\bf 0703} (2007) 066
  [arXiv:hep-ph/0612168].

\bibitem{gubserqhat}
  S.~S.~Gubser,
  ``Momentum fluctuations of heavy quarks in the gauge-string duality,''
  Nucl.\ Phys.\  B {\bf 790} (2008) 175
  [arXiv:hep-th/0612143].

\bibitem{ctqhat}
   J.~Casalderrey-Solana and D.~Teaney,
  ``Transverse momentum broadening of a fast quark in a $\cN = 4$ Yang Mills
  plasma,''
  JHEP {\bf 0704} (2007) 039
  [arXiv:hep-th/0701123].

\bibitem{argyres2}
  P.~C.~Argyres, M.~Edalati and J.~F.~V\'azquez-Poritz,
  ``Spacelike strings and jet quenching from a Wilson loop,''
  JHEP {\bf 0704} (2007) 049
  [arXiv:hep-th/0612157].

\bibitem{argyres3}
   P.~C.~Argyres, M.~Edalati and J.~F.~Vazquez-Poritz,
  ``Lightlike Wilson loops from AdS/CFT,''
  JHEP {\bf 0803}, 071 (2008)
  [arXiv:0801.4594 [hep-th]].

\bibitem{elr}
  F.~D'Eramo, H.~Liu and K.~Rajagopal,
  ``Transverse Momentum Broadening and the Jet Quenching Parameter, Redux,''
  arXiv:1006.1367 [hep-ph].

\bibitem{peigne1}
  S.~Peigne, P.~B.~Gossiaux and T.~Gousset,
  ``Retardation effect for collisional energy loss of hard partons produced  in
  a QGP,''
  JHEP {\bf 0604} (2006) 011
  [arXiv:hep-ph/0509185].

\bibitem{lorentzdirac}
  M.~Chernicoff, J.~A.~Garc\'\i a and A.~G\"uijosa,
  ``Generalized Lorentz-Dirac Equation for a Strongly-Coupled Gauge Theory,''
 Phys.\ Rev.\ Lett.\  {\bf 102} (2009) 241601
  [arXiv:0903.2047 [hep-th]].

  \bibitem{damping}
  M.~Chernicoff, J.~A.~Garc\'\i a and A.~G\"uijosa,
  ``A Tail of a Quark in $\cN=4$ SYM,''
  JHEP {\bf 0909} (2009) 080
  [arXiv:0906.1592 [hep-th]].

\bibitem{dragtime}
  M.~Chernicoff and A.~G\"uijosa,
  ``Acceleration, Energy Loss and Screening in Strongly-Coupled Gauge
  Theories,''
  JHEP {\bf 0806}, 005 (2008)
  [arXiv:0803.3070 [hep-th]].

\bibitem{dominguez}
  F.~Dominguez, C.~Marquet, A.~H.~Mueller, B.~Wu and B.~W.~Xiao,
  ``Comparing energy loss and $p_{\perp}$-broadening in perturbative QCD with
  strong coupling $\mathcal{N}=4$ SYM theory,''
  Nucl.\ Phys.\  A {\bf 811} (2008) 197
  [arXiv:0803.3234 [nucl-th]].

\bibitem{xiao}
  B.~W.~Xiao,
  ``On the exact solution of the accelerating string in $AdS_5$ space,''
  Phys.\ Lett.\  B {\bf 665} (2008) 173
  [arXiv:0804.1343 [hep-th]].

\bibitem{rangamani}
  J.~de Boer, V.~E.~Hubeny, M.~Rangamani and M.~Shigemori,
  ``Brownian motion in AdS/CFT,''
  JHEP {\bf 0907}, 094 (2009)
  [arXiv:0812.5112 [hep-th]].

\bibitem{sonteaney}
  D.~T.~Son and D.~Teaney,
  ``Thermal Noise and Stochastic Strings in AdS/CFT,''
  JHEP {\bf 0907}, 021 (2009)
  [arXiv:0901.2338 [hep-th]].

\bibitem{maldaeternal}
  J.~M.~Maldacena,
  ``Eternal black holes in Anti-de-Sitter,''
  JHEP {\bf 0304}, 021 (2003)
  [arXiv:hep-th/0106112].

\bibitem{herzogson}
  C.~P.~Herzog and D.~T.~Son,
  ``Schwinger-Keldysh propagators from AdS/CFT correspondence,''
  JHEP {\bf 0303}, 046 (2003)
  [arXiv:hep-th/0212072].

  \bibitem{iancu}
  G.~C.~Giecold, E.~Iancu and A.~H.~Mueller,
  ``Stochastic trailing string and Langevin dynamics from AdS/CFT,''
  JHEP {\bf 0907}, 033 (2009)
  [arXiv:0903.1840 [hep-th]].

\bibitem{giecold}
  G.~C.~Giecold,
  ``Heavy quark in an expanding plasma in AdS/CFT,''
  JHEP {\bf 0906} (2009) 002
  [arXiv:0904.1874 [hep-th]].

\bibitem{casalderrey}
  J.~Casalderrey-Solana, K.~Y.~Kim and D.~Teaney,
  ``Stochastic String Motion Above and Below the World Sheet Horizon,''
  JHEP {\bf 0912}, 066 (2009)
  [arXiv:0908.1470 [hep-th]].

  \bibitem{deboer}
  A.~N.~Atmaja, J.~de Boer and M.~Shigemori,
  ``Holographic Brownian Motion and Time Scales in Strongly Coupled Plasmas,''
  arXiv:1002.2429 [hep-th].

\bibitem{argyres1}
  P.~C.~Argyres, M.~Edalati and J.~F.~V\'azquez-Poritz,
  ``No-drag string configurations for steadily moving quark-antiquark pairs in
  a thermal bath,''
  JHEP {\bf 0701} (2007) 105
  [arXiv:hep-th/0608118].

\bibitem{mateosthermo}
  D.~Mateos, R.~C.~Myers and R.~M.~Thomson,
  ``Thermodynamics of the brane,''
  JHEP {\bf 0705}, 067 (2007)
  [arXiv:hep-th/0701132].

\bibitem{liu4}
   Q.~J.~Ejaz, T.~Faulkner, H.~Liu, K.~Rajagopal and U.~A.~Wiedemann,
  ``A limiting velocity for quarkonium propagation in a strongly coupled plasma
  via AdS/CFT,''
  JHEP {\bf 0804} (2008) 089
  [arXiv:0712.0590 [hep-th]].

\bibitem{mikhailov}
  A.~Mikhailov,
  ``Nonlinear waves in AdS/CFT correspondence,''
  arXiv:hep-th/0305196.

  \bibitem{uvir}
  L.~Susskind and E.~Witten,
  ``The Holographic Bound In Anti-De Sitter Space,''
  arXiv:hep-th/9805114;\\
  A.~W.~Peet and J.~Polchinski,
  ``UV/IR relations in AdS dynamics,''
  Phys.\ Rev.\ D {\bf 59} (1999) 065011
  [arXiv:hep-th/9809022].

\bibitem{fiol}
  B.~Fiol, B.~Garolera,
  ``Energy loss by radiation to all orders in $1/N$,''
  [arXiv:1106.5418 [hep-th]].

\bibitem{dirac}
  P.~A.~M.~Dirac,
  ``Classical theory of radiating electrons,''
  Proc.\ Roy.\ Soc.\ Lond.\  A {\bf 167} (1938) 148.

   \bibitem{teitelboim}
  C.~Teitelboim,
  ``Splitting of the Maxwell tensor --- radiation reaction without advanced
  fields,''
  Phys.\ Rev.\  D {\bf 1} (1970) 1572
  [Erratum-ibid.\  D {\bf 2} (1970) 1763].

\bibitem{rohrlich}
  F.~Rohrlich,
 \emph{Classical Charged Particles}, 2nd. ed. (Addison Wesley, Redwood City, California, 1990); 
  ``The dynamics of a charged sphere and the electron,''
  Am.\ J.\  Phys. {\bf 65} (1997) 1051.

  \bibitem{syz}
  E.~Shuryak, H.~-U.~Yee, I.~Zahed,
  ``Jet Quenching via Gravitational Radiation in Thermal AdS,''
  [arXiv:1110.0825 [hep-th]].

\bibitem{beuf}
  G.~Beuf, C.~Marquet and B.~W.~Xiao,
  ``Heavy-quark energy loss and thermalization in a strongly coupled SYM
  plasma,''
   Phys.\ Rev.\  D {\bf 80} (2009) 085001
  [arXiv:0812.1051 [hep-ph]].

\bibitem{eric}
Eric~J.~Pulido,
``P\'erdida de Energ\'{\i}a de Quarks en $\cN=4$ SYM,''
Master's Thesis, Universidad Nacional Aut\'onoma de M\'exico, in preparation.

   \bibitem{mo}
  K.~Maeda and T.~Okamura,
  ``Radiation from an accelerated quark in AdS/CFT,''
  arXiv:0712.4120 [hep-th].

\bibitem{fluct}
  H.~Dorn and H.~J.~Otto,
  ``Q anti-Q potential from AdS-CFT relation at $T \ge 0$: Dependence on
  orientation in internal space and higher curvature corrections,''
  JHEP {\bf 9809} (1998) 021
  [arXiv:hep-th/9807093];\\
  J.~Greensite and P.~Olesen,
  ``Worldsheet fluctuations and the heavy quark potential in the AdS/CFT
  approach,''
  JHEP {\bf 9904} (1999) 001
  [arXiv:hep-th/9901057].

  \bibitem{johnson2}
  P.~Johnson,
  ``Relativistic Particle Trajectories from Worldline Path Integral
  Quantization,''
{\it In the Proceedings of IEEE Particle Accelerator Conference (PAC 2001), Chicago, Illinois, 18-22 Jun 2001, pp 1781-1783}.

 \bibitem{brownian}
  E.~C\'aceres, M.~Chernicoff, A.~G\"uijosa and J.~F.~Pedraza,
  ``Quantum Fluctuations and the Unruh Effect in Strongly-Coupled Conformal
  Field Theories,''
  JHEP {\bf 1006} (2010) 078
  [arXiv:1003.5332 [hep-th]].

\bibitem{ppz}
  A.~Paredes, K.~Peeters and M.~Zamaklar,
  ``Temperature versus acceleration: the Unruh effect for holographic models,''
  JHEP {\bf 0904}, 015 (2009)
  [arXiv:0812.0981 [hep-th]].

  \bibitem{unruh}
  W.~G.~Unruh,
   ``Notes on black hole evaporation,''
   Phys.\ Rev.\  D {\bf 14} (1976) 870;\\
  P.~C.~W.~Davies,
  ``Scalar particle production in Schwarzschild and Rindler metrics,''
  J.\ Phys.\ A  {\bf 8}, 609 (1975).

 \bibitem{marianoangel}
  T.~Hirayama, P.~W.~Kao, S.~Kawamoto and F.~L.~Lin,
  ``Unruh effect and Holography,''
  arXiv:1001.1289 [hep-th];\\
   S.~R.~Das, T.~Nishioka, T.~Takayanagi,
  ``Probe Branes, Time-dependent Couplings and Thermalization in AdS/CFT,''
  JHEP {\bf 1007 } (2010)  071.
  [arXiv:1005.3348 [hep-th]];\\
     K.~Ghoroku, M.~Ishihara, K.~Kubo, T.~Taminato,
  ``Accelerated Quark and Holography for Confining Gauge theory,''
  Phys.\ Rev.\  {\bf D83 } (2011)  024020.
  [arXiv:1010.4396 [hep-th]];\\
  M.~Chernicoff, A.~Paredes,
  ``Accelerated detectors and worldsheet horizons in AdS/CFT,''
  JHEP {\bf 1103 } (2011)  063.
  [arXiv:1011.4206 [hep-th]].

  \bibitem{liustirring}
  K.~B.~Fadafan, H.~Liu, K.~Rajagopal and U.~A.~Wiedemann,
  ``Stirring Strongly Coupled Plasma,''
  Eur.\ Phys.\ J.\  C {\bf 61}, 553 (2009)
  [arXiv:0809.2869 [hep-ph]].

\bibitem{dampingtemp}
  A.~G\"uijosa and J.~F.~Pedraza,
  ``Early-Time Energy Loss in a Strongly-Coupled SYM Plasma,''
  JHEP {\bf 1105} (2011) 108
  [arXiv:1102.4893 [hep-th]].

\bibitem{sfetsosqqbar}
  S.~D.~Avramis, K.~Sfetsos and D.~Zoakos,
  ``On the velocity and chemical-potential dependence of the heavy-quark
  interaction in N = 4 SYM plasmas,''
  Phys.\ Rev.\  D {\bf 75} (2007) 025009
  [arXiv:hep-th/0609079].





\end{thebibliography}
\end{document}